\documentclass[a4paper,fleqn]{cas-dc}

\usepackage[numbers]{natbib}
\usepackage{graphicx}
\usepackage{amssymb}
\usepackage{longtable}
\usepackage{multirow}
\usepackage{appendix}
\usepackage[T1]{fontenc}
\usepackage{lscape}
\usepackage{amsmath}
\usepackage{caption} 
\usepackage{subcaption} 
\usepackage{booktabs}
\usepackage{tabularx}
\usepackage{algorithm2e}
\usepackage{vcell}
\usepackage{hhline}
\usepackage{balance}
\usepackage{float}
\usepackage{hyperref}
\usepackage{tikz}
\usepackage{url}
\usepackage{ulem}
\usepackage{graphicx}

\def\tsc#1{\csdef{#1}{\textsc{\lowercase{#1}}\xspace}}
\tsc{WGM}
\tsc{QE}
\tsc{EP}
\tsc{PMS}
\tsc{BEC}
\tsc{DE}

\begin{document}
\let\WriteBookmarks\relax
\def\floatpagepagefraction{1}
\def\textpagefraction{.001}

\shorttitle{}

\shortauthors{}


\title [mode = title]{Adaptive Data Quality Scoring Operations Framework using Drift-Aware Mechanism for Industrial Applications} 



%
\author[1]{Firas Bayram}[orcid=0000-0003-0683-2783]



\ead{firas.bayram@kau.se}


\affiliation[1]{organization={Department of Mathematics and Computer Science},
    addressline={}, 
    city={Karlstad},
    postcode={651 88}, 
    state={},
    country={Sweden}}

\author[1,2]{Bestoun S. Ahmed}[orcid=0000-0001-9051-7609]
\ead{bestoun@kau.se}

\affiliation[2]{organization={Department of Computer Science, Faculty of Electrical Engineering, Czech Technical University},
    city={Prague},
    country={Czech Republic}}

\author[3]{Erik Hallin}
\ead{erik.hallin@uddeholm.com}


\affiliation[3]{organization={Uddeholms AB},
    addressline={}, 
    city={Hagfors},
    postcode={683 33}, 
    state={V{\"a}rmlands l{\"a}n},
    country={Sweden}}





\begin{abstract}
Within data-driven artificial intelligence (AI) systems for industrial applications, ensuring the reliability of the incoming data streams is an integral part of trustworthy decision-making. An approach to assess data validity is data quality scoring, which assigns a score to each data point or stream based on various quality dimensions. However, certain dimensions exhibit dynamic qualities, which require adaptation on the basis of the system's current conditions. Existing methods often overlook this aspect, making them inefficient in dynamic production environments. In this paper, we introduce the Adaptive Data Quality Scoring Operations Framework, a novel framework developed to address the challenges posed by dynamic quality dimensions in industrial data streams. The framework introduces an innovative approach by integrating a dynamic change detector mechanism that actively monitors and adapts to changes in data quality, ensuring the relevance of quality scores. We evaluate the proposed framework performance in a real-world industrial use case. The experimental results reveal high predictive performance and efficient processing time, highlighting its effectiveness in practical quality-driven AI applications.

\end{abstract}



\begin{keywords}
Industrial application \sep Adaptive data quality \sep Data quality scoring \sep Data validation \sep Data-Driven AI \sep Drift detection
\end{keywords}

\maketitle

\section{Introduction}\label{sec:introduction}

Industries and businesses are actively accumulating data in unprecedented volumes, marking a definitive shift towards a data-centric paradigms \cite{reis2021data}. Within this landscape, data has transcended its conventional role to become the cornerstone of success for artificial intelligence (AI) software solutions. Data holds an intrinsic value due to its inseparable link to the life cycle of machine learning (ML), which constitutes the primary type of AI. Therefore, assessing the quality of the collected data becomes an essential and imperative aspect in building robust and reliable ML solutions \cite{hazen2014data}. Remarkably, a prevalent concern in the industry is the disproportionate allocation of efforts in ML projects in research. According to a recent study conducted by MIT scientists \cite{Stonebraker2019MachineLA}, research institutions often allocate 90\% of their efforts to improve ML algorithms and only 10\% to data preparation and validation. The authors suggested that these numbers should be reversed for better overall outcomes. In light of this, the significance of data quality (DQ) assurance extends beyond mere procedural correctness. It plays a pivotal role in influencing the cost-effectiveness and operational efficiency of industrial processes and operations.


There are two main approaches to data quality assessment: quantitative assessments that involve scales and metrics to quantify aspects of data quality, and qualitative assessments that focus on inherent characteristics and subjective evaluations \cite{zaveri2016quality}. Quantitative assessments provide a measurement, offering more detailed information about the objective estimates of data quality. This approach involves assigning specific numerical values to various aspects, called a quality score or index \cite{chen2014review}, which enables a quantifiable understanding of data quality. In contrast, qualitative assessments contribute to a more holistic perspective by exploring the intrinsic qualities and subjective aspects of the data, referred to as data profiling \cite{liu2009encyclopedia}. Through the use of scales and metrics, our study focuses on quantitative data scoring, as it provides valuable insights for industries.

Data quality scoring is a methodological approach that involves evaluating and assigning scores to data records based on predefined criteria across several data quality dimensions \cite{batini2009methodologies}. Each data quality dimension captures a unique aspect, collectively contributing to a thorough assessment of the overall data quality. The scores indicate the acceptance level of the addressed data quality dimensions, distinguishing between high- and poor-quality data. Furthermore, data quality scores are interpreted as the degree to which data quality conforms to specified aspects of data quality \cite{taleb2021big}. In the context of large-scale systems, the importance of these data quality scores extends beyond the evaluation of data quality and can be exploited to enhance the overall performance of the ML system \cite{chen2021data}.

From a data-centric AI perspective, assessing data quality is crucial; nevertheless, it adds computational costs to systems, especially in real-time production environments. The practicality of these assessments in operational contexts poses a notable challenge due to inherent complexities and resource-intensive nature in industrial use cases. Therefore, streamlining these processes becomes a vital requirement to ensure their effectiveness and success in practical applications \cite{budach2022effects}. Two primary approaches exist for data quality scoring: a traditional standard approach and an automated approach utilizing an ML-based system \cite{widad2023quality}. The standard approach involves checking a predefined set of data quality dimensions and assigning scores to each dimension. In contrast, ML-based methods employ an ML regressor to predict the score of the processed data window instead of the manual scoring procedure.

In the domain of industrial processes where real-time data analysis is crucial, ML-based methods present benefits in terms of efficiency and speed \cite{wu2019real}. Using ML algorithms can significantly reduce the time and resources required for data quality assessment, especially in large-scale environments \cite{bayram2023dqsops}. The choice of an ML-based approach over traditional methods is driven by the need for a more scalable, efficient, and adaptive solution. ML-based methods can process large volumes of data quickly in real-time, providing a more timely evaluation of data quality. The predictive capabilities of ML models enable faster analysis and decision-making, enhancing the overall effectiveness of data quality evaluation processes. This capability is particularly beneficial in industries such as manufacturing, where early identification and resolution of data quality issues can optimize production processes and minimize downtime.

ML-based approaches can be categorized into adaptive and non-adaptive methods \cite{clerc2016adaptive}. Regarding data quality assessment, adaptivity is not only about ML model adjustments. Rather, it refers to the ability to respond effectively to dynamic changes in data quality dimensions. These adaptive methods continuously monitor and update their quality assessment criteria based on incoming data, allowing them to respond effectively to evolving conditions. In contrast, non-adaptive methods rely on static quality assessment criteria throughout the analysis, potentially overlooking changes in data quality over time.

Existing data quality scoring frameworks often neglect the crucial aspect of adaptivity, presenting a major obstacle to effective management of dynamic data quality. Specifically, non-adaptive ML frameworks have two inherent limitations that need to be addressed. Firstly, determining the optimal size of the optimal time checkpoint to retrain the ML model in production can be challenging and may lead to inefficiencies. Secondly, certain data quality dimensions exhibit a dynamic nature, reflecting the fluctuating conditions of industrial systems. For example, highly relevant data in one phase may be regarded as of a lower quality under different circumstances. The limitations of non-adaptive frameworks highlight the importance of retraining the ML model to capture these dynamic changes. ML models are trained on historical data to learn the characteristics of high- and low-quality data according to current conditions. When data characteristics change, quality scores also change, and an adaptation signal should be triggered to adapt the ML model to the evolving data quality characteristics, ensuring constant learning for more accurate assessments.

This paper introduces an innovative approach to tackle the dynamic challenges of data quality assessment by proposing a novel framework that integrates adaptivity into ML-based data quality scoring methodologies. Our main contribution lies in addressing the evolving nature of data quality and incorporating drift detection mechanisms to enhance data quality scoring accuracy. The proposed framework dynamically adjusts the retraining process based on evolving data patterns, enabling a more precise and adaptable data quality assessment over time. In addition, we introduce adaptive mechanisms to facilitate dynamic responses and re-calibration of data quality scores according to the prevailing system conditions. This approach effectively meets the demands of large-scale industrial processes where data quality requirements evolve continuously.

The remainder of the paper is structured as follows. Section \ref{sec:background} outlines the conceptual background necessary to understand our proposed framework. In Section \ref{sec:related}, we provide an overview of the existing literature related to data quality assessments. Following this, Section \ref{sec:adqsops} introduces the framework with its development and deployment phases. Section \ref{sec:results} presents implementation details along with an analysis of its predictive performance and processing time efficiency. Finally, in Section \ref{sec:conclusion}, we summarize the key contributions of the paper and discuss potential opportunities for future research.

\section{Conceptual Background}\label{sec:background}
\begin{figure}
    \centering
    \includegraphics[width=1\linewidth]{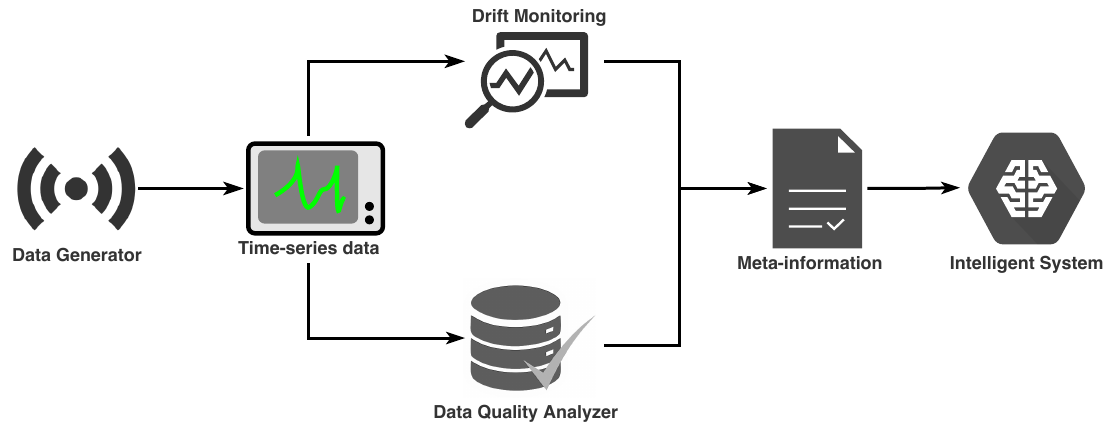}
    \caption{Quality-driven data workflow for informed decision-making in industrial applications.}
    \label{fig:workflow}
\end{figure}

Continuous monitoring and assessment of the incoming data streams are crucial to ensure that dynamic changes are captured and that the prevalent condition (concept) remains consistent. In data-centric applications, it is essential to verify that data quality remains high. Data quality is important in ensuring the data remains useful, as low-quality data can lead to physical failures or inaccuracies in sensor readings \cite{teh2020sensor}. This process is illustrated in Figure \ref{fig:workflow}, where time-series data collected from the data source undergoes both drift detection and data quality assessment. Consequently, the AI system benefits from valuable meta-information obtained from these stages. Continuous drift monitoring ensures that the incoming data remain contextually similar to the training data \cite{polyzotis2018data}, allowing intelligent systems to make informed decisions based on high-quality information \cite{zha2023data}.

Within the context of data-driven applications, these changes can have implications for the performance of both data quality and application-level ML systems. They signify the need for adaptation to maintain system reliability. In practice, application-level ML systems utilize drift detection to sustain performance in dynamic environments, while data-quality scoring ML systems use it to ensure the quality and reliability of the generated data. Despite their different objectives, both application-level and data quality scoring ML systems rely on the same core principle: monitoring changes in data characteristics. This common foundation allows for the successful application of identical drift detection mechanisms across both model types. This paper focuses primarily on adapting data quality scoring ML systems within the holistic AI application. This section provides essential background knowledge on these core concepts, laying the foundation for a comprehensive understanding of the fundamental elements within the scope of this research.

\subsection{Drift Monitoring}\label{sec:drift_detection}
Drift in data streams refers to the temporal evolution or evolving changes in the underlying statistical properties of the data \cite{agrahari2022concept}. Detecting drift involves employing a methodology to determine and identify significant changes in data streams at a specific time point $t$. In industrial settings, detecting drift is a fundamental task that serves to indicate and diagnose the status of the ingested data, prompting appropriate actions in response to identified changes \cite{ditzler2015learning}. Typically, drift is detected by quantifying the dissimilarity between the data probability distributions of two timestamps using the divergence metric. If the divergence metric exceeds a certain threshold $\zeta$, it signifies a drift occurred in the data \cite{liu2013change}. This metric can be calculated as follows:
\begin{equation}
D_{t} = \delta(P \parallel Q).
\end{equation}
The decision-making logic is defined as follows: If $D_{t} > \zeta$, a drift in the data is detected; otherwise, no detection occurs. Where \(D_{t}\) represents the divergence metric recorded at timestamp $t$ between probability distributions \(P\) and \(Q\), and \(\delta\) is a function measuring this divergence.

In the context of nonstationary applications, determining a suitable threshold value $\zeta$ for detecting drift poses a significant challenge \cite{micevska2021sddm}. Industrial data streams are usually dynamic and complex, resulting in constantly changing statistical properties. This variability makes establishing a single, universal threshold impractical. Moreover, effectively detecting drift requires a subtle understanding of the specific industrial processes and the context of the application \cite{kammerer2019anomaly}. As a result, recent research concludes that the drift detection threshold should not be fixed. Instead, it must be adaptive, adjusting dynamically to reflect the evolving nature of the system and its underlying processes \cite{liu2022concept}.

To overcome these challenges in developing our proposed adaptive data quality scoring framework, we have integrated a dynamic method from our prior research, designed to efficiently detect changes in time-series data distributions \cite{bayram2023lstm}. This method has been Incorporated into our proposed framework to monitor the distribution of incoming application data. Specifically, it actively observes the divergence values computed for each time window frame using a sliding window mechanism. The divergence value is calculated using the Jensen-Shannon divergence, defined as follows \cite{lin1991divergence}:

\begin{equation}
\label{entropy}
\mathrm{JSD}(P_{\text{his}} \parallel P_{\text{cur}}):= H\left(\frac{p_{\text{his}}+p_{\text{cur}}}{2}\right)-\frac{H(p_{\text{his}})+H(p_{\text{cur}})}{2},
\end{equation}
where $P_{\text{his}}$ and $P_{\text{cur}}$ represent the probability distributions for the historical and current application data samples, respectively, and the function $H$ is the Shannon's entropy given by:
\begin{equation}
H(p)=-\int p(\mathbf{Y}) \log p(\mathbf{Y}) d \mathbf{Y}.
\end{equation}


Subsequently, the algorithm makes decisions on drift detection by performing hypothesis testing on the p-value of the observed divergence value, indicating the extremeness of the current change magnitude with respect to historical magnitudes. If the p-value falls below the defined significance threshold $\tau$, a drift detection signal is triggered. An inherent advantage of this method lies in its avoidance of the requirement of a predefined threshold for drift magnitude. Furthermore, its dynamic nature ensures robustness and adaptability as the divergence distribution evolves with the accumulation of more data, making it a viable solution for the evolving nature of real-world industrial applications.


\subsection{Data Quality Assurance}

In the industrial applications of ML systems, the assurance of data quality is fundamental to developing high-performance decision-making support \cite{wang2023overview}. In particular, rigorous data quality practices are essential to ensure that the data used for both training and inference are of optimal quality and represent the underlying processes in a timely manner \cite{fan2022foundations}. Low-quality data in industrial processes that are driven by data analytics could be caused by machine errors, inconsistent sensor measurements, or abnormal patterns, among other interpretations. Therefore, data quality assurance is recognized as a critical factor that significantly impacts the cost-effectiveness and operational efficiency of industrial processes, addressing potential faults and improving overall performance.

Data quality assurance is a multi-dimensional concept that spans various attributes to collectively assess the validity of collected data \cite{pipino2002data}. Each dimension offers a unique perspective on the specific characteristics of the data. Numerous studies have compiled extensive lists, some identifying up to 179 distinct quality dimensions \cite{wang1996beyond}. These dimensions are further grouped into \textit{intrinsic}, \textit{contextual}, \textit{accessibility}, and \textit{representational} categories \cite{priestley2023survey}. The selection of data quality dimensions is not uniform, as there is no universally accepted definition. Rather, they are often defined based on the specific goals and requirements of a given task and application \cite{karkouch2016data}.

Based on the data context of our use case, we have identified and selected specific data quality dimensions that are highly relevant and applicable to the characteristics of our industrial application. We summarize the definitions of data quality dimensions as follows:

\begin{enumerate}
    \item \textbf{Accuracy}: This dimension evaluates how well the recorded data aligns with the actual values it is meant to represent. It evaluates precision and correctness, which are essential for detecting anomalies that could lead to product defects.

    \item \textbf{Completeness}: This dimension assesses the thoroughness of the observed data by checking for missing values that are collected from the data source. Ensuring that all relevant sensor data is captured and no sensor failures occur.

    \item \textbf{Consistency}: This dimension examines whether the observed values conform to the integrity constraints of the domain, ensuring that collected data records fall within the expected value ranges. This contributes to effective product quality monitoring and prevents errors due to incorrect sensor readings.

    \item \textbf{Timeliness}: Describing the relevance of data for specific tasks, timeliness assesses whether the observed data is current. In dynamic industrial settings, where ongoing tasks often demand up-to-date information, ensuring the currency of collected data becomes of high importance. This ensures that data is still representing the system and can be trusted for the task.

    \item \textbf{Skewness}: This dimension goes beyond traditional measures, computing the distribution deviation of observed data from a reference distribution. Addressing skewness is important for optimized performance, especially for ML systems, as skewed data distributions can affect model accuracy and generalization. It verifies that no unexpected changes or unusual patterns exist.
\end{enumerate}

Observing the defined data quality dimensions, a key point of consideration emerges within the context of the ML systems: a subset of these dimensions, including timeliness and skewness, exhibit \textit{dynamic characteristics}. This implies that the quality aspects of these dimensions may vary according to the prevailing conditions. To illustrate, for the timeliness dimension, what is deemed timely and well-fitted in the present may not remain true in the future. Similarly, the skewness of the data may vary with different seasons, introducing variables such as drift and seasonality. Therefore, data demonstrating drift during a specific season, resulting in a low skewness score for the current season, might conversely indicate a high score in a different season. On the other hand, certain dimensions, such as accuracy, completeness, and consistency, exhibit \textit{constant characteristics}, maintaining their relevance irrespective of changing conditions. These constant dimensions focus on the precision and validity of data representation rather than relying on transient factors and settings within the problem domain. Therefore, the integration of the ML-based scoring framework with an adaptation methodology specifically addresses the adaptive nature of data quality in dynamic dimensions within data-centric ML applications. This integration specifically addresses the adaptive nature of data quality in data-centric ML applications.

\section{Related Work}\label{sec:related}

The pivotal role of data quality assessment in ensuring the reliability and effectiveness of data-driven processes has been examined in diverse applications and domains, highlighting its growing recognition in the research community. In the field of healthcare, such as electronic health records (EHRs), researchers have explored methodologies to assess and improve the quality of patient data, acknowledging its critical impact on medical decision-making and patient care \cite{lewis2023electronic}. Within the financial sector, studies have focused on assessing the quality of financial data to maintain the integrity of analytical models and regulatory compliance \cite{karkovskova2023data, hasan2020current}. Assessment of spatial data quality has been studied by introducing a comprehensive quality assessment framework for linear features of Volunteered Geographic Information (VGI) by integrating novel quality metrics with those commonly used through factor analysis \cite{wu2021comprehensive}. The common thread between these diverse applications is the recognition of data quality as a foundational element for robust and trustworthy results \cite{mcgilvray2021executing}.

The use of IoT technologies in industries to gather and generate data from IoT sensors, often in real-time, requires a rigorous evaluation of data quality \cite{mansouri2023iot}. Taleb et al. \cite{taleb2021big} presented the Big Data Quality Management Framework (BDQMF) as an exhaustive strategy aimed at addressing data quality challenges inherent in large-scale data systems. The framework defines various dimensions of data quality and incorporates multiple components dedicated to managing, validating, and monitoring data quality. It includes a scoring mechanism to quantify different aspects of data quality. The authors explored issues related to data quality at both the individual cell instance and the schema levels within datasets. Another recent approach in this context is the big data quality assessment framework (BIGQA) \cite{fadlallah2023bigqa}. The framework provides a declarative solution specifically designed for non-expert users, featuring reporting functionality to visualize outcomes or scores indicating the quality of input datasets. In a different approach, the Data Quality Anomaly Detection Framework focuses on anomaly detection \cite{widad2023quality}, based on an extended isolation forest model. It introduces the \textit{Quality Anomaly Score} metric to evaluate the degree of anomalousness in six dimensions of quality. 

In addition to these studies, Chug \textit{et al.} \cite{chug2021statistical} have developed a method to assess dataset quality using nine dimensions, yielding a \textit{comprehensive score}, \textit{report}, and \textit{label}. The study introduced \textit{data quality ingredients} as semantic indicators, identifying nine crucial aspects, including provenance, characteristics, uniformity, metadata coupling, missing cells, duplicate rows, data skewness, inconsistency ratio in categorical columns, and attribute correlation. These ingredients contribute to the calculation of the final score. In a previous study, Ardagna \textit{et al.} \cite{ardagna2018context} proposed a method to enhance computational efficiency by focusing on a specific data subset, reducing both time and resource requirements using parallelization. To convey the reliability of data quality values, they introduced a \textit{confidence} metric tied to the considered data volume and influenced by time constraints and computational resources. The confidence metric serves as an indicator of the trustworthiness of the data in relation to the quality dimension of the data tested. 

Another recent research Byabazaire \textit{et al.} \cite{byabazaire2022end} has explored \textit{trust} metrics as a means of assessing real-time data quality in IoT deployments. The proposed framework allows end-users to tailor trust metrics, offering visibility into data quality throughout the big data model. Building on this trust metric for real-time data quality assessment in IoT deployments, another paper \cite{byabazaire2023iot} extends the previous approach to data quality assessment. This study introduces a solution that employs data fusion strategies, specifically Adaptive Weighted Fusion, Kalman’s Fusion, and Naïve Fusion, to derive a unified quality score. The practical experiments aim to evaluate the computational efficiency of these different fusion methods. 

The analysis of the existing literature highlights the widespread use of traditional methods in evaluating data quality, revealing a lack of transformative frameworks and significant breakthroughs. Furthermore, the majority of studies tend to assess complete datasets, which could pose challenges to ML systems. A more effective approach lies in the real-time scoring of the data streams, which offers a balance between efficiency and granularity that better suits the dynamic nature of ML systems. This aligns with the inherent reliance of ML systems on individual data records, as scoring at the record level allows for a more granular assessment of data quality. In addition, the significant costs associated with traditional scoring methods require alternative approaches. Therefore, several recent studies have explored ML-based techniques as a promising solution to mitigate these drawbacks. One such method is presented by Widad \textit{et al.} \cite{widad2023quality}, which utilizes an intelligent anomaly detection model to score data quality anomalies using an unsupervised ML model. The data quality anomaly scoring framework detects anomalies in multiple quality dimensions and quantifies the extent of deviation from established quality standards. In another research, our previous framework Data Quality Scoring Operations (DQSOps) \cite{bayram2023dqsops}, based on DevOps principles, specifically continuous integration/continuous delivery (CI/CD), enables efficient real-time data quality evaluation in industrial applications. The framework scores several data quality dimensions and demonstrated significantly faster processing rates compared to traditional scoring processes. Although these ML-based methods effectively score the quality of data streams in a production environment, they lack an adaptive nature for data quality assessment, which we address in the current research.



\section{The Adaptive Data Quality Scoring Operations Framework}\label{sec:adqsops}
\begin{figure*}
    \centering
    \includegraphics[width=1\linewidth]{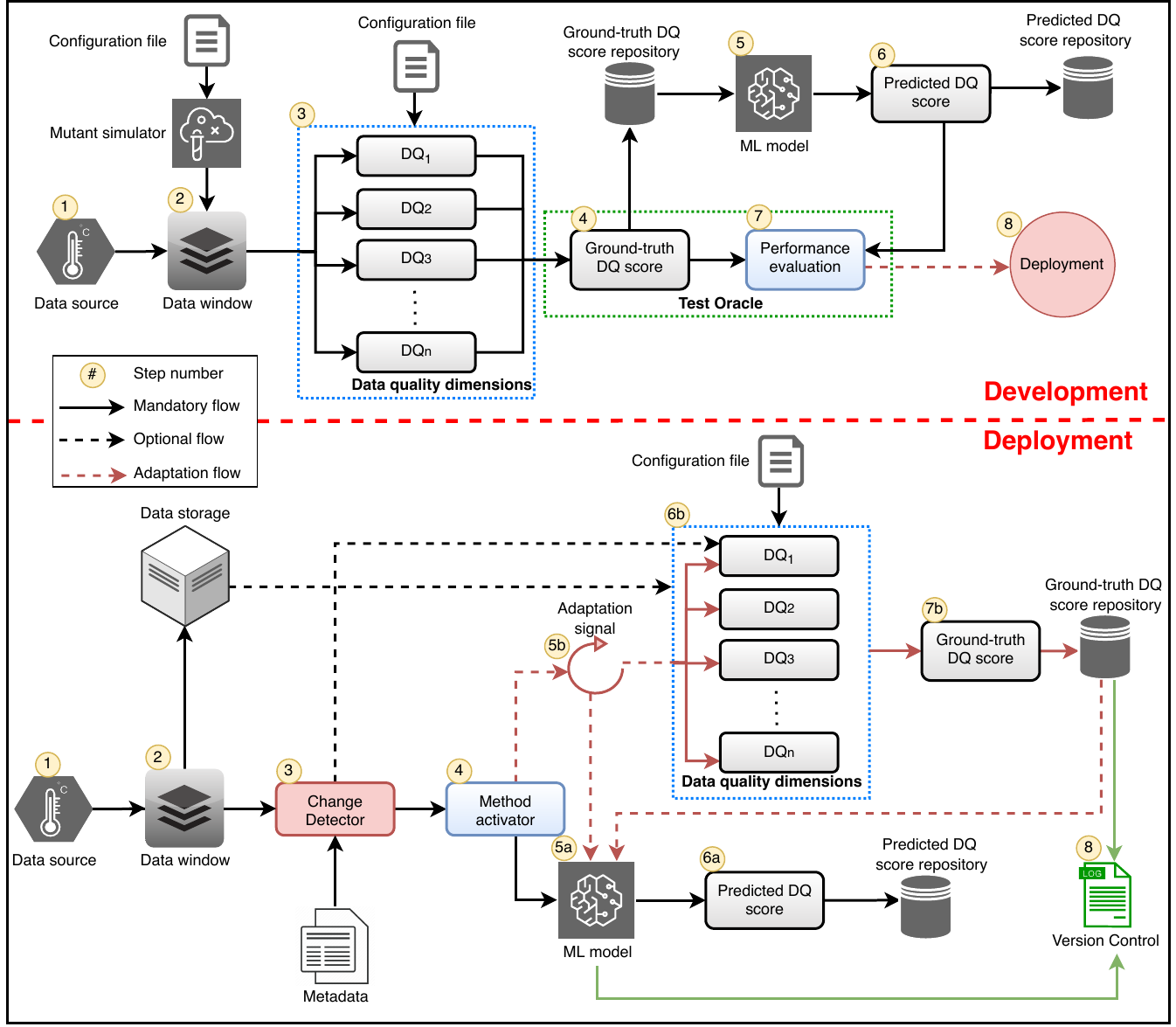}
    \caption{The Adaptive Data Quality Scoring Operations framework.}
    \label{fig:A-DQSOps}
\end{figure*}

The proposed adaptive data quality scoring framework employs an ML-based approach to score the data quality in industrial applications. The primary objective of the framework is to label the incoming data windows with a score that quantifies the quality of the data window based on various pre-defined data quality dimensions. This innovative framework has been designed to address the intrinsic limitations of the non-adaptive static data quality scoring framework, especially in industrial contexts characterized by dynamic data environments. Specifically, the novel adaptive framework effectively handles the dynamic aspects of data quality by re-assessing the data quality scores based on the prevalent conditions observed in the application system. This adaptation mechanism, which is based on the drift detection method, allows the dynamic data quality dimensions to conform to the evolving characteristics of the underlying data and ML processes. Furthermore, the drift detection method eliminates the need to define a fixed window size $w$ to retrain the ML model, since the retraining signal is activated only when a drift is detected. The workflow of the proposed adaptive data quality scoring framework is illustrated in Figure \ref{fig:A-DQSOps}.

First, we provide an overview of the underlying ML-based scoring framework to establish a foundation and enhance understanding. This overview summarizes the framework's core components, specifically highlighting its implementation of MLOps practices to manage the ML regressor for data quality scoring. This context is crucial to understanding the operational environment of the proposed framework. Subsequently, we dissect the two integral phases to achieve the proposed framework: development and deployment phases. The development phase focuses on initializing the framework and generating the necessary artifacts. These artifacts are then streamlined in the deployment phase, enabling dynamic scoring of incoming data windows based on their quality.

\subsection{An Overview of ML-Based Data Quality Scoring}\label{sec:dqsop}
The ML-based data quality scoring approach aims to determine the data quality of the collected data window to generate a unified score using an ML model \cite{evans2006scaling}. This score reflects the overall adequacy of the data windows, considering multiple dimensions of data quality. Specifically, the ML-based scoring framework is designed to address the complexity of scoring data quality in data-driven applications  \cite{singh2015quality}. The main innovation in the framework is streamlining the scoring process by using an ML predictor instead of traditional standard scoring methods following MLOps principles, incorporating continuous monitoring and validation practices. This integration provides significant speedup rates while maintaining high predictive performance levels. Additionally, the framework's runtime remains unaffected by the number of quality dimensions, offering practical scalability in real-world applications with high sampling rates. 

To accomplish ML-based scoring operations, the framework workflow begins by initializing the ML predictor in a warm-start mode. In particular, the ML predictor is initiated using training data that contain ground-truth quality scores obtained by the standard-based approach. The model's prediction accuracy is monitored using a \textit{test oracle} until it reaches a predetermined threshold $T$ based on a performance metric. Once this threshold is met, the ML predictor is eventually deployed in the real-world problem. Meta-information files, such as the anomaly detection model and reference data distribution, are also prepared to calculate certain data quality dimension scores. Furthermore, as part of this phase, a mutant simulator is integrated to improve the model's learning by introducing a variety of data quality issues that may arise in practical situations, thus accelerating the learning process. 

Once the framework is deployed in a production environment, the method activator becomes a key component in managing the pipeline flow. It employs predetermined criteria to select the appropriate approach to data quality scoring. Specifically, the activator chooses between the ML-based approach and the standard-based approach based on the collected data window and chunk size. The activator repeatedly executes the ML model to obtain data quality scores until the chunk size reaches a preconfigured threshold $\beta$. At this point, an evaluation is initiated using the standard-based approach to ensure continuous monitoring of the ML model's accuracy. This evaluation employs a specified test oracle to compare predicted quality scores with ground-truth scores, and if the model's performance falls below a pre-defined tolerance level, a retrain signal is activated. 

The process of finding the consolidated data quality score involves aggregating the calculated values of the data quality dimensions. Traditional methods like the arithmetic mean may not be suitable due to their sensitivity to variable scales. To address this, the quality scores are standardized using z-scores, ensuring uniform integration of different data quality metrics \cite{heinrich2018requirements}. This standardization involves calculating the z-score for each element in the quality score matrix. Principal Component Analysis (PCA) is then employed, following widely used techniques in the literature \cite{teh2020sensor}. 

PCA condenses the information from multiple dimensions into a single interpretable score, simplifying the analysis and facilitating easier comparison of data quality across different data windows or even datasets. Specifically, the first principal component, which indicates the direction of maximum variance, is used as a score that signifies the overall quality of the data windows. This approach allows us to summarize the essential information from multiple dimensions into a single comprehensive score. This ensures that the overall quality assessment is more intuitive and easily understandable, facilitating better interpretation. Moreover, using a single comprehensive score rather than multiple dimensions makes the assessment process simpler and more straightforward for decision-making in our industrial context.

The quality of each time-series data is evaluated across several key dimensions, each representing a distinct aspect of data quality. Let \(\text{TS}_i\) represent a time series, where \(i\) denotes the index of the time series. The quality of each time series \(\text{TS}_i\) is evaluated across several key dimensions \(\text{DQ}_1, \text{DQ}_2, \dots, \text{DQ}_n\), each representing a specific aspect of data quality. The evaluation process yields a corresponding data quality score \(\text{DQS}_i\) for the time series \(\text{TS}_i\), which is computed as follows:
\[
\text{TS}_i \rightarrow \text{PCA}(\text{DQ}_1, \dots, \text{DQ}_n) = \text{DQS}_i.
\] For the calculation of the individual data quality scores, the ground-truth labels for each dimension are determined using commonly used criteria in the literature \cite{taleb2021big, bayram2023dqsops}.:

\begin{enumerate}
    \item \textbf{Accuracy Score:} Calculated as the proportion of anomalous data in the window: \(Accuracy = \frac{NAV}{N}\), where \(NAV\) is total anomalous values, and \(N\) is the data window size.

    \item \textbf{Completeness Score:} Quantifies missing values:\\ \(Completeness = \frac{NNV}{N}\), where \(NNV\) is the number of missing values (NA or NULL), and \(N\) is the window size.
    
    \item \textbf{Consistency Score:} Calculated based on defined integrity constraints for data values: \(Consistency = \frac{NCV}{N}\), where \(NCV\) is the number of consistent values, and \(N\) is the window size.
    
    \item \textbf{Timeliness Score:} Involves a goodness-of-fit test, particularly a two-sample Kolmogorov-Smirnov test, comparing current data with an expected distribution. The Kolmogorov-Smirnov test statistic is computed as \(KS = \max_{1 \leq i \leq N}|\hat{F_1}(Z_i) - \hat{F_2}(Z_i)|\), where \(Z\) is the combined sample of two independent random samples \(X\) and \(Y\).
    
    \item \textbf{Skewness Score:} Utilizes the Jensen-Shannon Divergence (JSD) value to measure dissimilarity between the distributions of current and historical data. JSD is calculated as \(\mathrm{JSD}(P\| Q) = H\left(\frac{P+Q}{2}\right) - \frac{H(P)+H(Q)}{2}\), where \(H\) denotes Shannon's entropy. JSD is chosen for its bounded nature in the interval \([0,1]\) \cite{lionis2021rssi}.
\end{enumerate}

\subsection{Development Phase}
Similar to any supervised ML system that requires a warm start for effective initialization and optimal performance \cite{ash2020warm}, the development phase of our proposed framework starts with the initiation of crucial system artifacts essential for streamlining the solution in production. This phase includes tasks such as system setup, parameter configuration, establishing the anomaly detector, defining data distribution building parameters, managing metadata to calculate data quality dimension scores, ML model development, and preparing all necessary components for subsequent deployment. Additionally, certain meta-information helps reduce the number of calculations in the deployment phase. Table \ref{tab:init_artifacts} summarizes the overall artifacts produced in the development phase. The overall workflow of this phase is illustrated in Figure \ref{fig:A-DQSOps}. The process begins with data collection from the data source \textbf{(Step 1)}, which is then segmented into data windows \textbf{(Step 2)}. These data windows are processed through various data quality dimensions \textbf{(Step 3)}, and the necessary auxiliary information for calculating these dimensions, such as integrity constraints for consistency score and file paths, is loaded from a configuration file. This information is used to calculate the individual ground-truth quality scores \textbf{(Step 4)}.

The core outcome of this phase is the creation of the ML predictor for data quality \textbf{(Step 5)}. This predictor is trained on historical data labeled with ground-truth quality scores \textbf{(Step 6)}. These scores are calculated using the standard approach detailed in Section \ref{sec:dqsop}. To enhance the predictor's learning capabilities and accelerate training, a \textit{data mutant simulator} is employed. This component introduces potential data quality issues that may occur in real-world scenarios, enhancing the predictor's ability to learn data quality issues and potentially reducing the amount of real-world data required. The mutation parameters are stored in configuration files, which include settings like the percentage of faults to be simulated. Continuous performance monitoring is implemented using a \textit{test oracle} to systematically evaluate the performance of the ML predictor \textbf{(Step 7)}, with a predefined threshold for a selected performance metric. Consistent achievement of this threshold indicates sufficient learning for deployment to the production environment \textbf{(Step 8)}, marking the completion of the development phase.

\begin{table}
    \centering
    \caption{Artifacts produced in the development phase}
    \begin{tabular}{|l|p{5cm}|}
        \hline
        \textbf{Artifact} & \textbf{Description} \\
        \hline
        \textbf{ML Predictor} & The ML model that is used to predict quality score. \\
        \hline
        \textbf{Divergence values} & Used for the dynamic drift detection method to detect changes. \\
        \hline
        \textbf{Anomaly detector} & Utilized in the calculation of the accuracy score. \\
        \hline
        \textbf{Data distribution} & Represents the historical PDF used to calculate skewness score. \\
        \hline
        \textbf{Historical samples} & Used in the timeliness score calculation. \\
        \hline
    \end{tabular}
    \label{tab:init_artifacts}
\end{table}

\subsection{Deployment Phase} \label{sec:deployment}
Moving from the development environment to the production environment, the deployment phase of our adaptive framework involves incorporating the developed and related components of the ML predictor and integrating them into the operational system. To manage the ML system in this phase, the principles of MLOps are followed, including CI/CD. The adoption of MLOps deployment strategies results in a systematic and efficient process for deploying and maintaining the ML model in a live operational context \cite{kreuzberger2023machine}. The MLOps ecosystem encompasses a broad set of practices, tools, and techniques designed to automate the ML life cycle. This includes version control, automated testing, continuous monitoring, and continuous model updates, which makes it valid for industrial applications with minimal intervention \cite{kumara2023requirements}.

\begin{figure}
    \centering
    \includegraphics[width=1\linewidth]{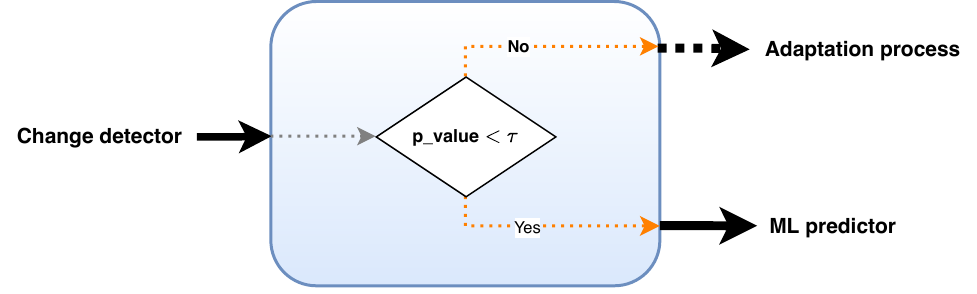}
    \caption{Method activator component flowchart.}
    \label{fig:method_activator}
\end{figure}

As illustrated in Figure \ref{fig:A-DQSOps}, the real-time data collected from the source \textbf{(Step 1)} is segmented into data windows \textbf{(Step 2)} during the deployment phase. Subsequently, a change detector is applied to the data window to assess the occurrence of drift \textbf{(Step 3)}. The change detector performs hypothesis testing to determine the significance of the drift magnitude, as explained in detail in Section \ref{sec:drift_detection}. The change detector leverages meta-information collected during the development phase, such as historical PDF for data and divergence values, both of which are used in making the calculations in this step. The result is then forwarded to the method activator component, which in turn makes decisions about which flow to proceed with \textbf{(Step 4)}. The flow chart diagram of the activator component of the method is depicted in Figure \ref{fig:method_activator}.

In the absence of detected drift, the method activator component continues using the current ML model \textbf{(Step 5a)} without initiating adaptation to make the DQ scoring predictions \textbf{(Step 6a)}. However, in the case of drift detection, the activator component of the method initiates a retraining signal \textbf{(Step 5b)}. This signal triggers the adaptation process flow, which simulates the re-scoring of the historical data based on the prevailing conditions and updates the scores of the dynamic data quality dimensions, and scoring the newly collected data \textbf{(Step 6b)}. This mechanism ensures that the previous ground-truth labels are properly updated to reflect the changes in the dynamic data quality dimensions. Subsequently, the ML model is retrained from scratch using both the updated development-time training data and the newly scored data points \textbf{(Step 7b)}. This approach ensures that the model fully integrates the updated ground-truth labels for historical data, reflecting the changes in dynamic data quality dimensions. Additionally, a version control system is employed to manage and track updates to both the data quality scores and the ML model \textbf{(Step 8)}, ensuring that frequent updates are systematically documented and managed.

\color{black}

\section{Experimental Results}\label{sec:results}
To evaluate our proposed framework, practical experiments were carried out in collaboration with Uddeholms AB, a leading steel manufacturer\footnote{https://www.uddeholm.com/en/}. These experiments were carried out to assess the efficiency of the framework from various perspectives, including the accuracy of predictive performance, execution time, and resource consumption over time. Furthermore, we conducted a comparative analysis between our adaptive approach, the static data scoring approach, and the standard scoring approach. This analysis is presented to provide detailed insight into the merits and limitations of each approach. The following subsection presents a detailed description of the industrial use case and implementation, followed by a thorough analysis and discussion of the experimental results.

\subsection{Implementation and Use Case Details}
\begin{figure}
    \centering
    \includegraphics[width=1\linewidth]{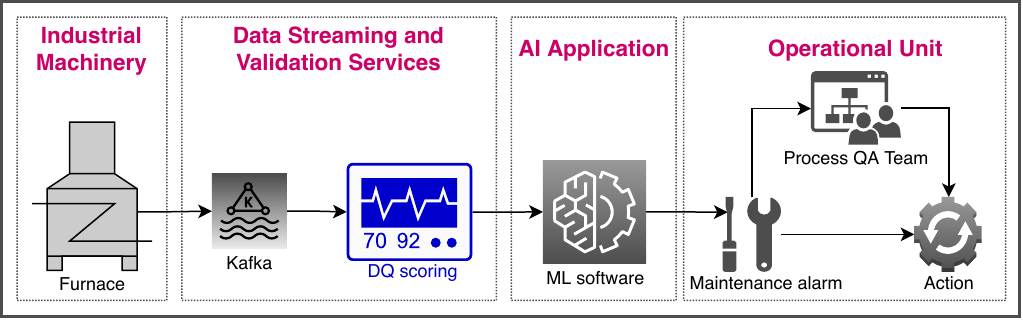}
    \caption{Integration of data quality scoring in the holistic industrial use case application.}
    \label{fig:usecase}
\end{figure}

Our adaptive approach is designed to be streamlined into a broader software system specifically developed to assist decision-making in industrial processes. In this set of experiments, the proposed framework was implemented and evaluated in the industrial application use case of the Electroslag Remelting (ESR) vacuum pumping process at the Uddeholm steel manufacturer in Sweden. The overall AI-driven application aims to sustain the production of high-quality steel by monitoring the pressure values. A representation view of our studied application is shown in Figure \ref{fig:usecase}.

Starting with industrial machinery, the figure shows the key ingredients of the application, including the furnace where a sensor collects pressure data within the vacuum chamber. Each time the vacuum pump is activated, which can take up to 20 minutes, pressure values are continuously monitored in this use case. The sensor records values every millisecond and through the Apache Kafka streaming platform\footnote{https://kafka.apache.org/}, data windows are transmitted every second for real-time analysis. Subsequently, each data window proceeds to data streaming and validation services, passing through a data quality scoring framework. Subsequently, the scoring framework validates the collected data, assesses its quality, and generates a score. Following this, in the AI-driven decision-making process, alarms are triggered for improper pump events. These alerts are communicated to the maintenance team of the operational unit, which takes appropriate actions, such as stopping the pump event to eliminate its costs. The primary purpose of the application is to achieve a gradual decrease in pressure during appropriate pump events, reaching the desired minimum value within the allocated time while promptly identifying and addressing improper pump events to prevent interruptions and uphold optimal furnace operation.

In terms of implementation details, our proposed framework was built with the Python programming language, taking advantage of its multipurpose capabilities and wide range of libraries. The implementation also incorporates the YAML format for defining configuration files. These YAML-formatted configuration files allow for the clear and organized specification of different parameters and settings, ultimately enabling the flexibility and customization of the framework to meet specific requirements of the use case. For the ML models, we employ the widely-used extreme gradient boosting (XGBoost) model \cite{chen2016xgboost}. This decision tree-based ensemble technique has proven effective across various applications, demonstrating its suitability for industrial contexts and its capability to provide strong and easily understandable results \cite{kiangala2021effective}.


\subsection{Drift Detection Sensitivity Analysis}
\begin{figure}
    \centering
    \includegraphics[width=1\linewidth]{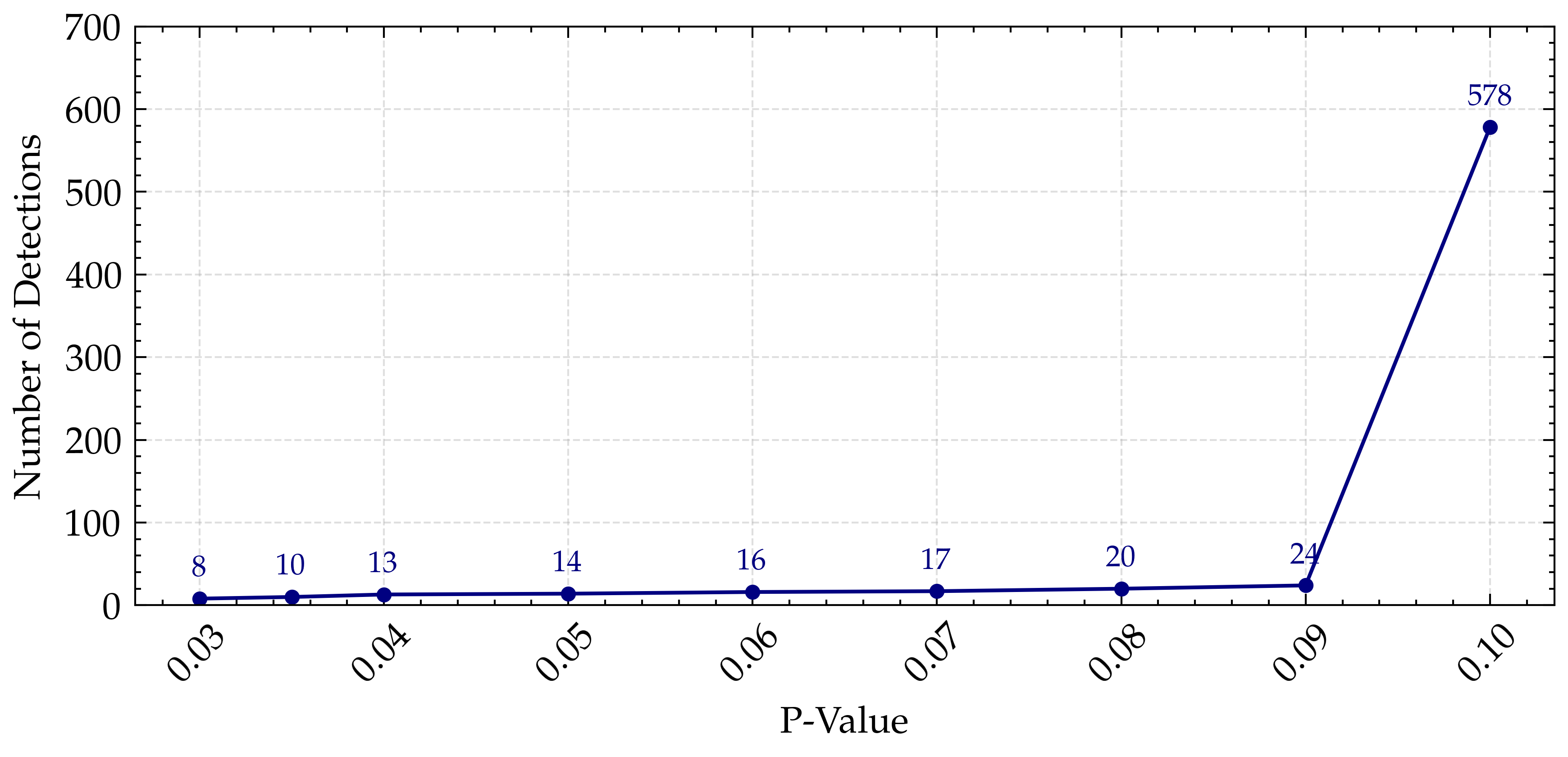}
    \caption{Changes detected at different p-values.}
    \label{fig:pvalue_changes}
\end{figure}

In a real-world production environment, the effectiveness of the adaptive data quality scoring approach is highly dependent on the change detection component to initiate the adaptation signal, a mechanism detailed in Section \ref{sec:deployment}. A sensitivity analysis of the drift detection mechanism in the approach was conducted to understand the behavior of the change detector. This analysis involved varying the significance thresholds represented by the p-values. The results, visually presented in Figure \ref{fig:pvalue_changes}, illustrate the number of changes detected at different p-values, with the values tested including $\tau= [0.03, 0.035, 0.04, 0.05, 0.06, 0.07, 0.08, 0.09, 0.1]$.

The analysis reveals a gradual increase in the number of detections from a significance level $\tau$ of 0.03 to 0.09, ranging from 8 detections to 24, respectively, demonstrating a steady response to subtle variations. However, a significant spike is observed at a p-value of 0.1, reaching 578 detections, indicating a potential higher sensitivity that could lead to false alarms. Therefore, to maintain a diverse set of detection numbers in our experiments, we selected p-values of 0.03, 0.04, 0.06, 0.08, and 0.09, excluding 0.1 due to its observed high sensitivity.

\subsection{Performance Analysis of DQ Scoring Predictions}
The predictive performance of our data quality scoring ML model in the adaptive framework through a series of experiments using the defined levels of significance threshold $\tau$. The evaluation involves calculating the errors of the XGBoost regressor results, measured by two metrics: Mean Absolute Error (MAE) and R-squared (R2), which are given by:\color{black}
\begin{equation}
\textit{MAE} = \frac{1}{n} \sum_{i=1}^{n} |y_i - \hat{y}_i|,
\end{equation}

\begin{equation}
R^2 = 1 - \frac{\sum_{i=1}^{n} (y_i - \hat{y}_i)^2}{\sum_{i=1}^{n} (y_i - \bar{y})^2}.
\end{equation}
Where $n$ is the number of data points, $y_i$ is the actual value, $\hat{y}_i$ is the predicted value, and $\bar{y}$ is the mean of the actual values.

\begin{figure*}
    \centering
    \begin{subfigure}[b]{0.48\textwidth}
        \centering
        \includegraphics[width=\linewidth]{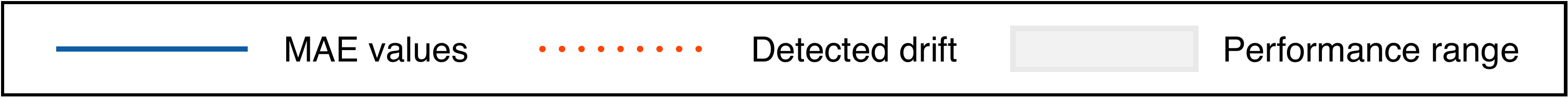}
    \end{subfigure}

    \begin{subfigure}[b]{0.48\textwidth}
        \centering
        \includegraphics[width=\linewidth]{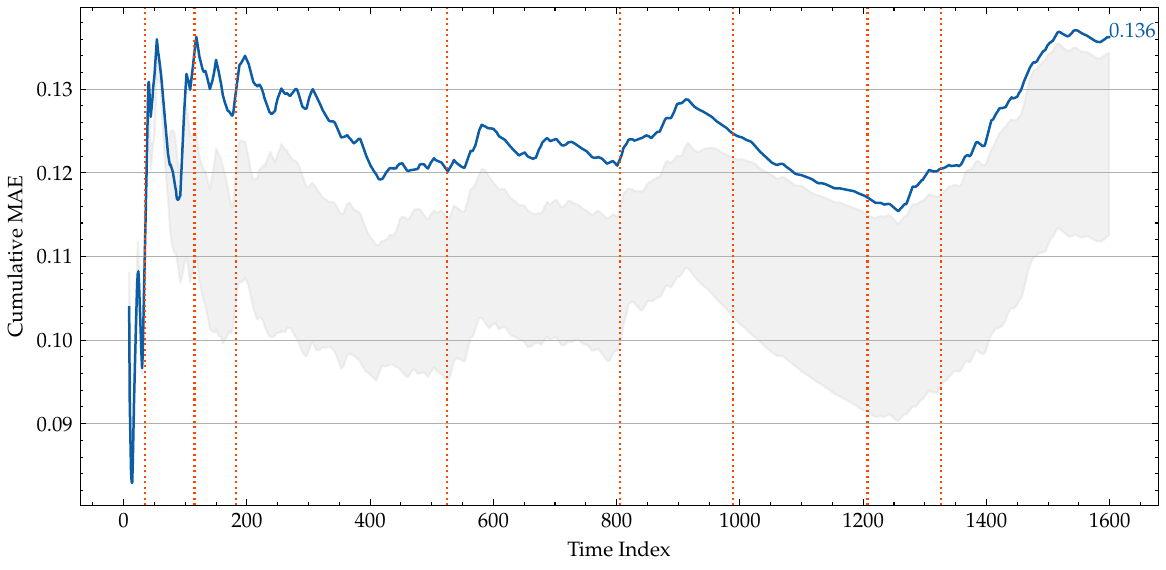}
        \caption{Adaptive approach with \( \tau = 0.03 \)}
    \end{subfigure}
    \hfill
    \begin{subfigure}[b]{0.48\textwidth}
        \centering
        \includegraphics[width=\linewidth]{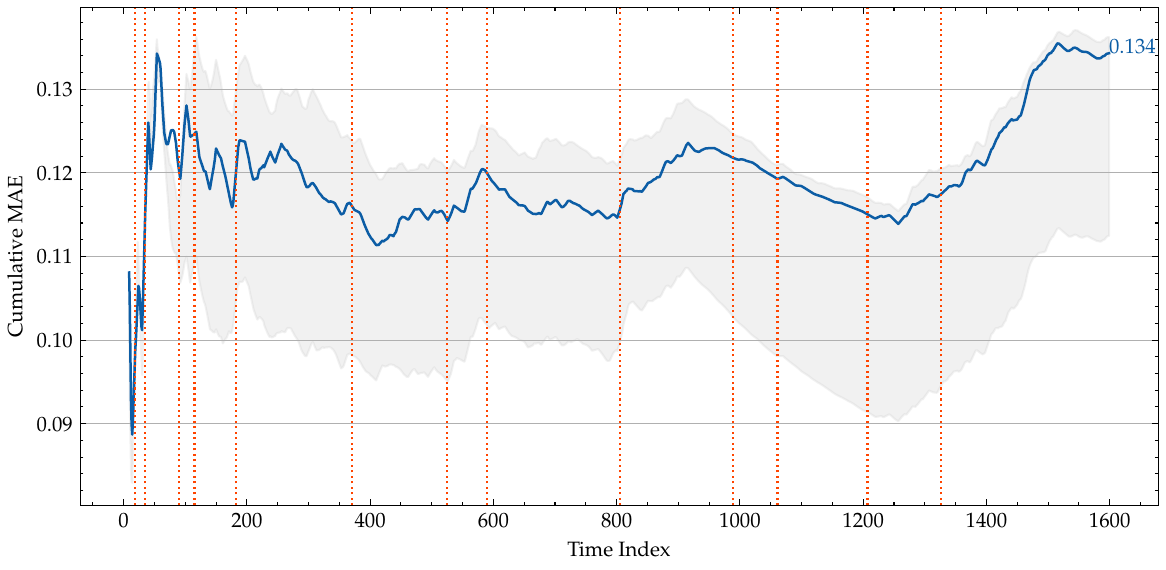}
        \caption{Adaptive approach with \( \tau = 0.04 \)}
    \end{subfigure}
    
        \vspace{5pt}

    \begin{subfigure}[b]{0.48\textwidth}
        \centering
        \includegraphics[width=\linewidth]{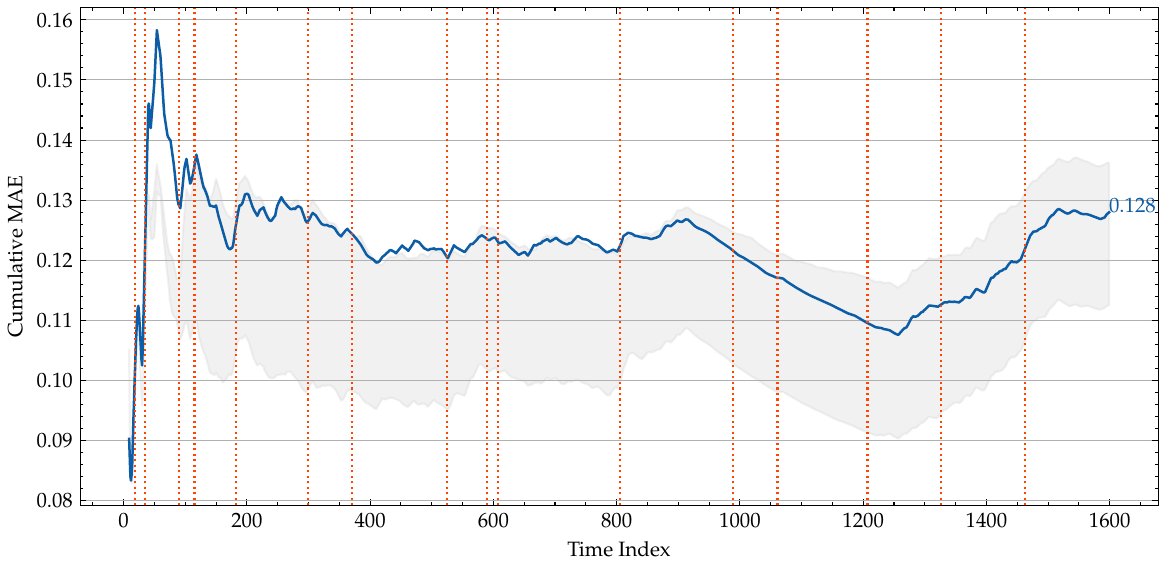}
        \caption{Adaptive approach with \( \tau = 0.06 \)}
    \end{subfigure}
    \hfill
    \begin{subfigure}[b]{0.48\textwidth}
        \centering
        \includegraphics[width=\linewidth]{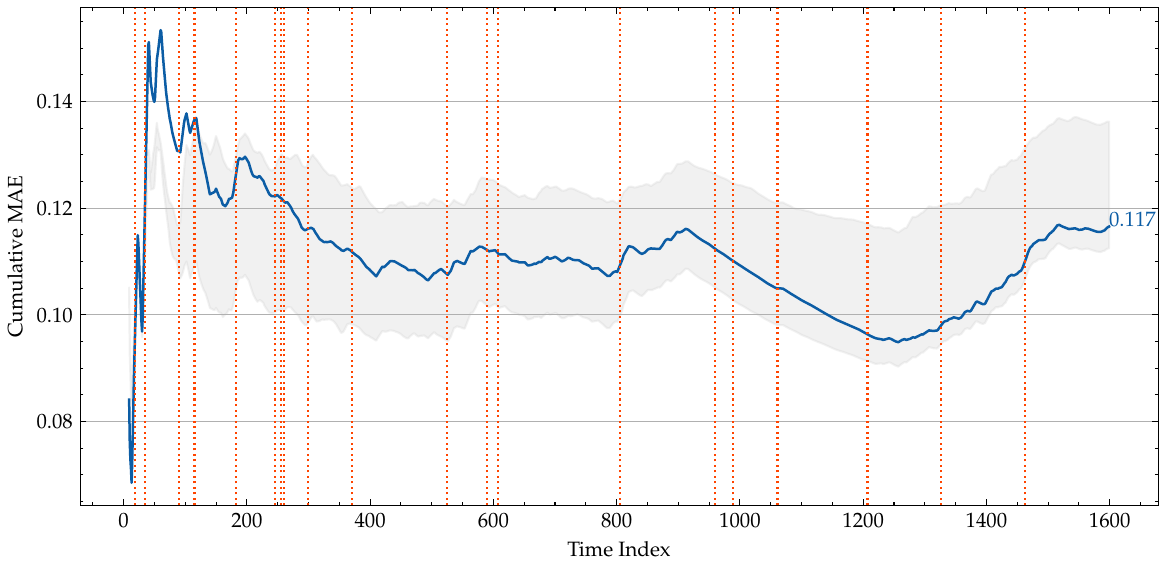}
        \caption{Adaptive approach with \( \tau = 0.08 \)}
    \end{subfigure}
    
    \vspace{5pt}
    
    \begin{subfigure}[b]{0.48\textwidth}
        \centering
        \includegraphics[width=\linewidth]{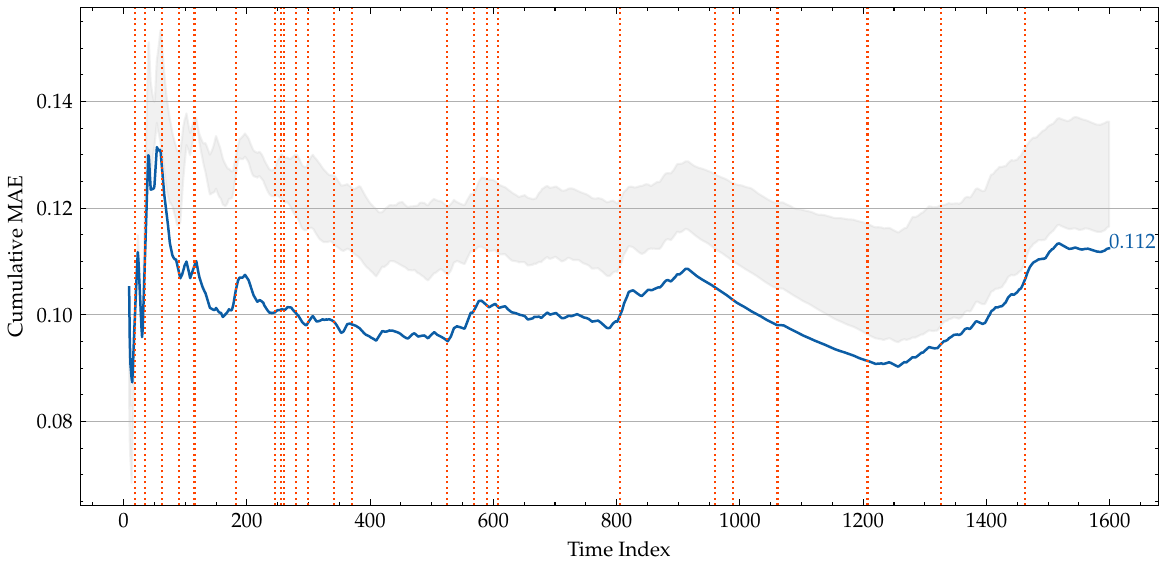}
        \caption{Adaptive approach with \( \tau = 0.09 \)}
    \end{subfigure}

    \caption{Cumulative MAE for our adaptive approach with varying levels of \( \tau \). Shaded area denotes performance range across other \(\tau\) values.}
    \label{fig:adqsop_mae}
\end{figure*}

\begin{figure*}
    \centering
    \begin{subfigure}[b]{0.48\textwidth}
        \centering
        \includegraphics[width=\linewidth]{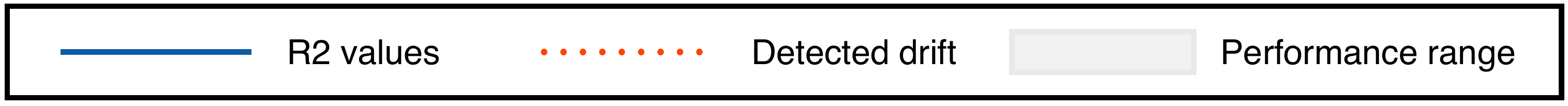}
    \end{subfigure}

    \begin{subfigure}[b]{0.48\textwidth}
        \centering
        \includegraphics[width=\linewidth]{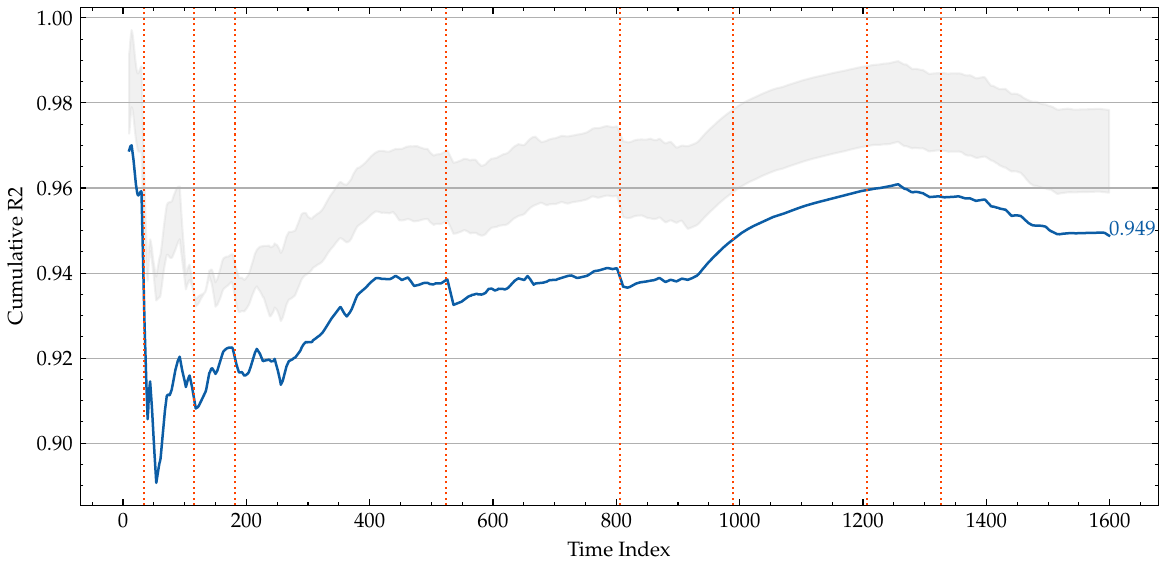}
        \caption{Adaptive approach with \( \tau = 0.03 \)}
    \end{subfigure}
    \hfill
    \begin{subfigure}[b]{0.48\textwidth}
        \centering
        \includegraphics[width=\linewidth]{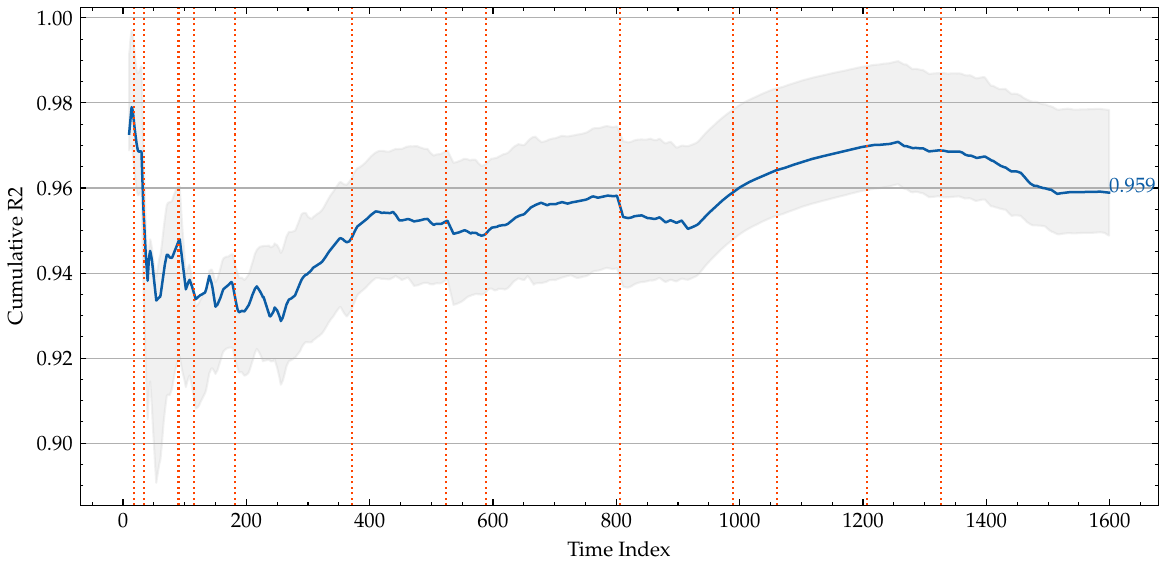}
        \caption{Adaptive approach with \( \tau = 0.04 \)}
    \end{subfigure}
    
    \vspace{5pt}  

    \begin{subfigure}[b]{0.48\textwidth}
        \centering
        \includegraphics[width=\linewidth]{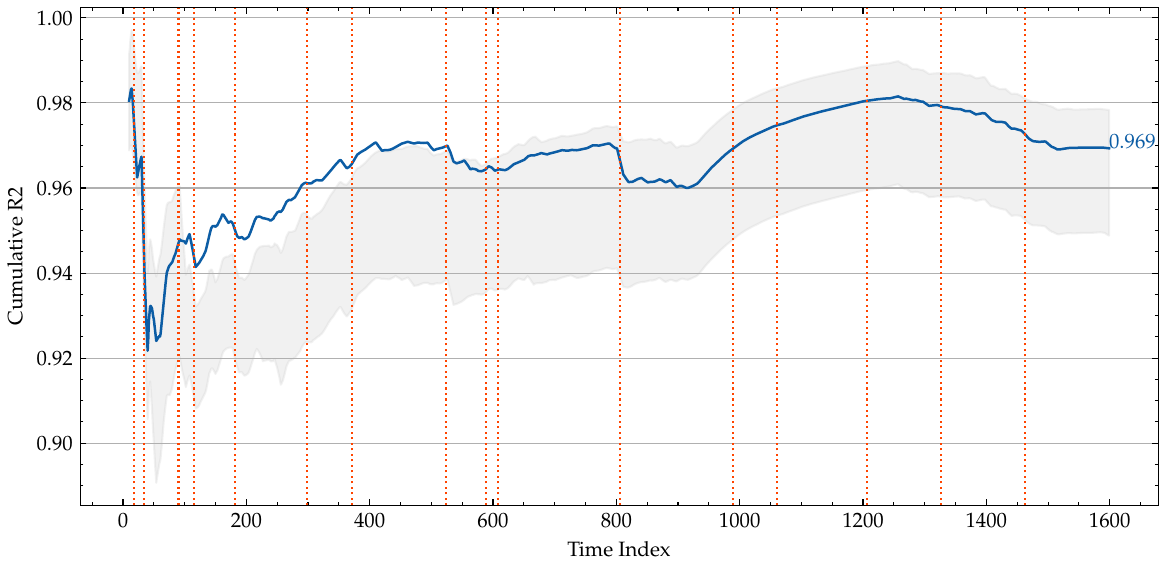}
        \caption{Adaptive approach with \( \tau = 0.06 \)}
    \end{subfigure}
    \hfill
    \begin{subfigure}[b]{0.48\textwidth}
        \centering
        \includegraphics[width=\linewidth]{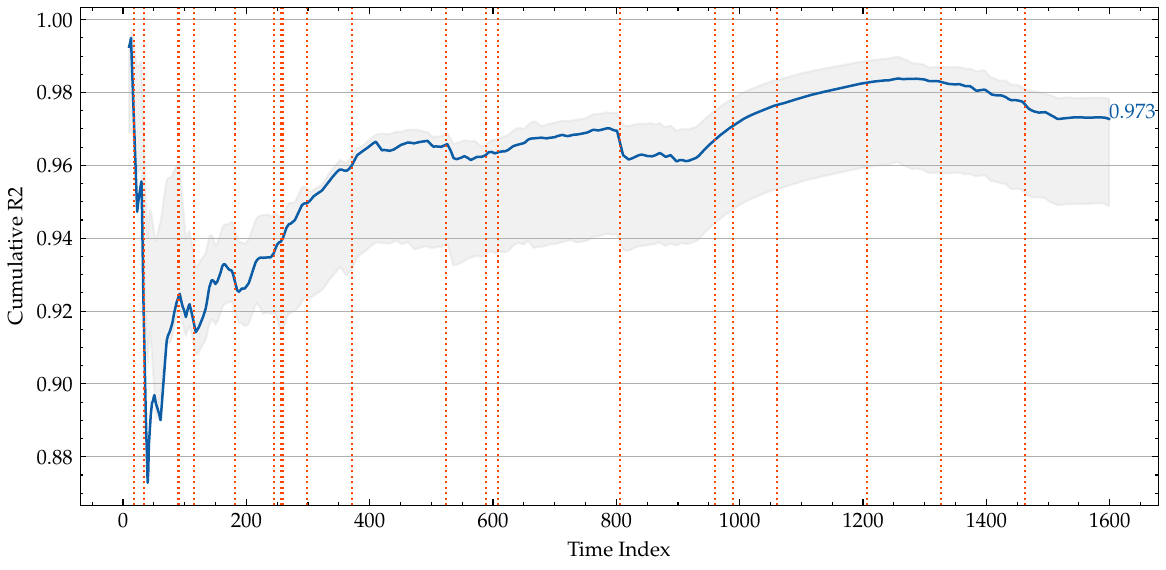}
        \caption{Adaptive approach with \( \tau = 0.08 \)}
    \end{subfigure}
    
    \vspace{5pt}  

    \begin{subfigure}[b]{0.48\textwidth}
        \centering
        \includegraphics[width=\linewidth]{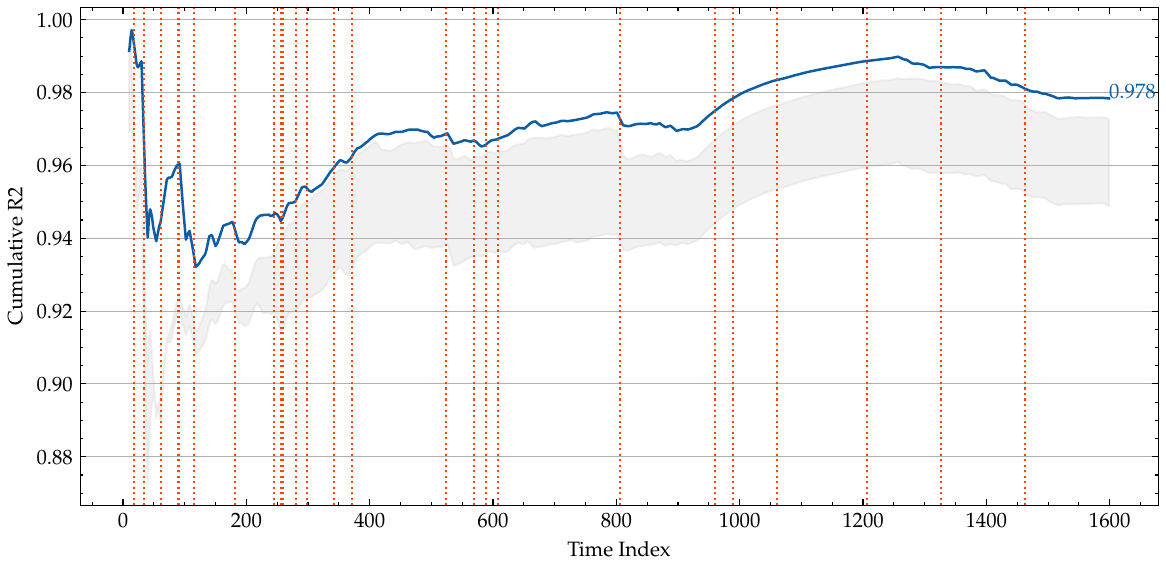}
        \caption{Adaptive approach with \( \tau = 0.09 \)}
    \end{subfigure}

    \caption{Cumulative R2 for our adaptive approach with varying levels of \( \tau \). Shaded area denotes the performance range across \(\tau\) values.}
    \label{fig:adqsop_r2}
\end{figure*}

The predictive performance of the framework was evaluated by examining the observed errors over time, specifically MAE, presented in Figure \ref{fig:adqsop_mae}, and R2 in Figure \ref{fig:adqsop_r2}. Each subfigure displays the error metric along with the corresponding detected drifts for each significance level $\tau$, annotated with shaded areas representing the highest and lowest performances among the remaining levels. The results show that drifts are detected; hence, adaptation is executed more frequently in the initial stages of deployment than in later stages. This observation suggests that the drift detection mechanism becomes more robust over time, resulting in fewer false alarms as more divergence values are collected.

For prediction errors, the results show that after the execution of adaptation mechanisms prompted by drift detection, the predictive performance often improves. This improvement is particularly reflected in the decline in MAE and an increase in $R^2$ metrics after adaptation. Specifically, at the end of the experimental duration, the MAE metric values are 0.136 for experiments with $\tau = 0.03$ and 0.112 for experiments with $\tau = 0.09$. The values for the $R^2$ metric are 0.949 for experiments with $\tau = 0.03$ and 0.978 for experiments with $\tau = 0.09$. These variations in results are attributed to the fact that a lower $\tau$ leads to a reduced sensitivity, resulting in fewer triggered adaptations that may not be sufficient to update the predictor, while a higher $\tau$ leads to more frequent adaptations.

\begin{figure*}
    \centering
    \begin{subfigure}[b]{0.4\textwidth}
        \centering
        \includegraphics[width=\linewidth]{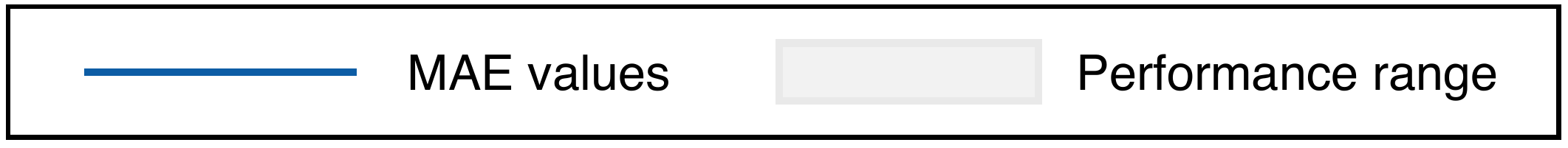}
    \end{subfigure}

    \begin{subfigure}[b]{0.48\textwidth}
        \centering
        \includegraphics[width=\linewidth]{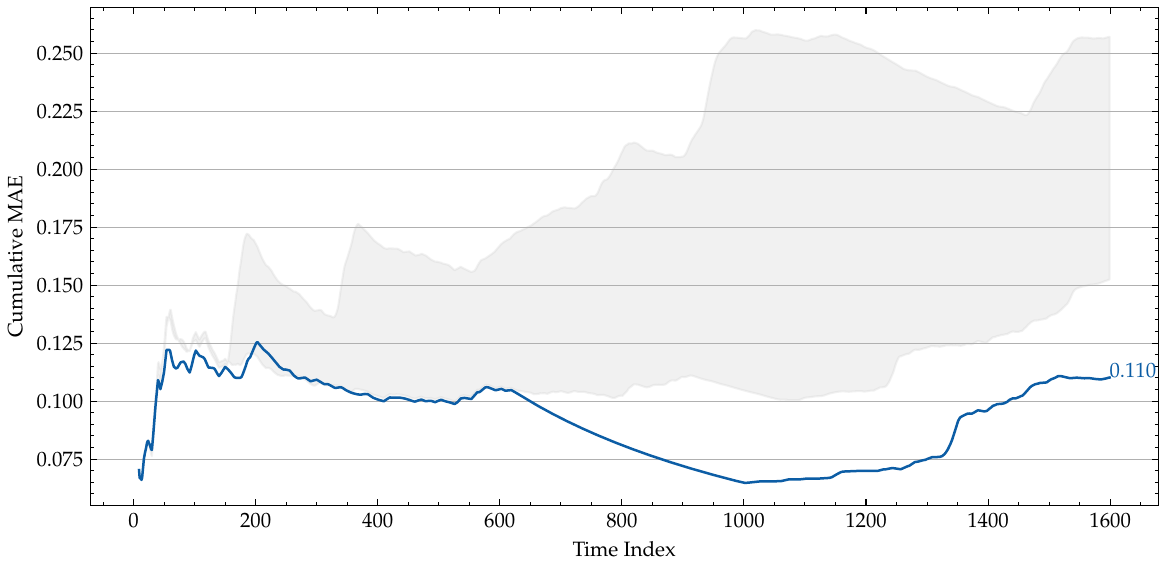}
        \caption{Static approach with $w=25$}
        \label{fig:mae_25}
    \end{subfigure}
    \hfill
    \begin{subfigure}[b]{0.48\textwidth}
        \centering
        \includegraphics[width=\linewidth]{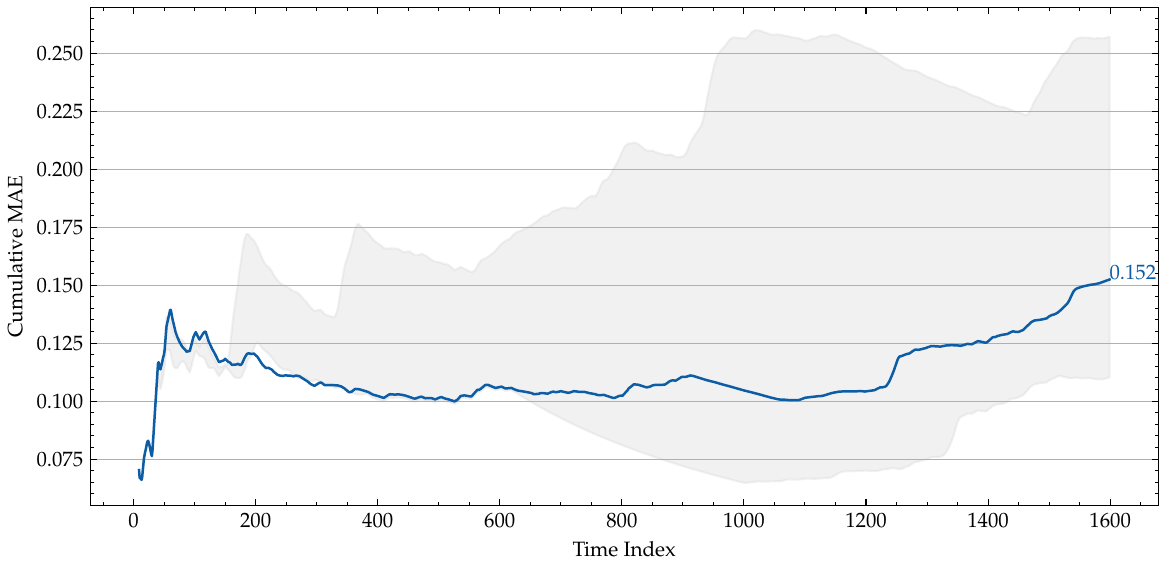}
        \caption{Static approach with $w=50$}
        \label{fig:mae_50}
    \end{subfigure}

    \vspace{10pt}

    \begin{subfigure}[b]{0.48\textwidth}
        \centering
        \includegraphics[width=\linewidth]{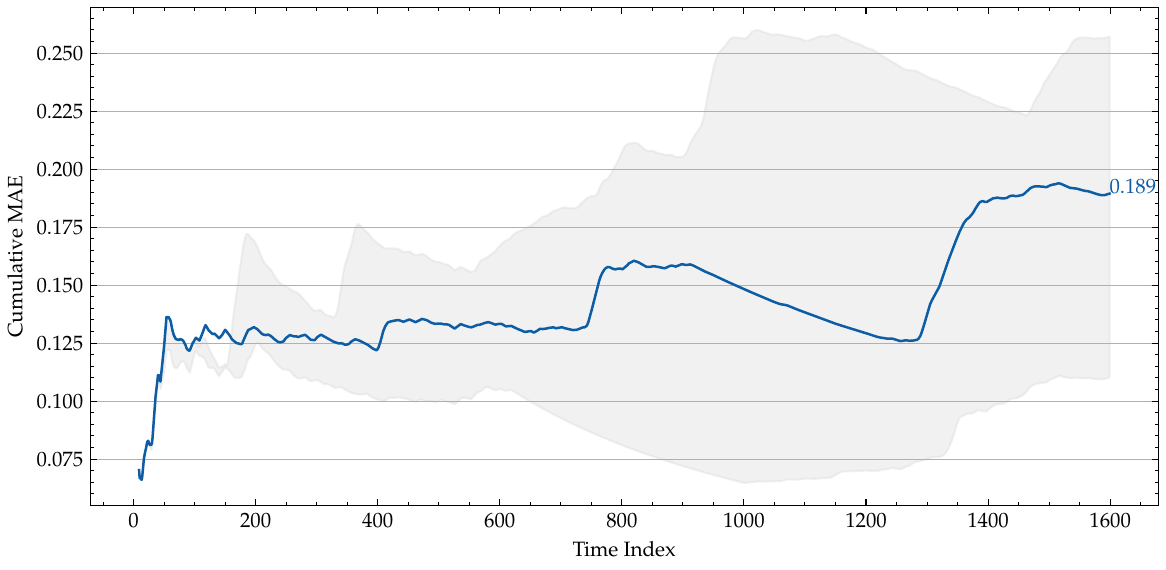}
        \caption{Static approach with $w=100$}
        \label{fig:mae_100}
    \end{subfigure}
    \hfill
    \begin{subfigure}[b]{0.48\textwidth}
        \centering
        \includegraphics[width=\linewidth]{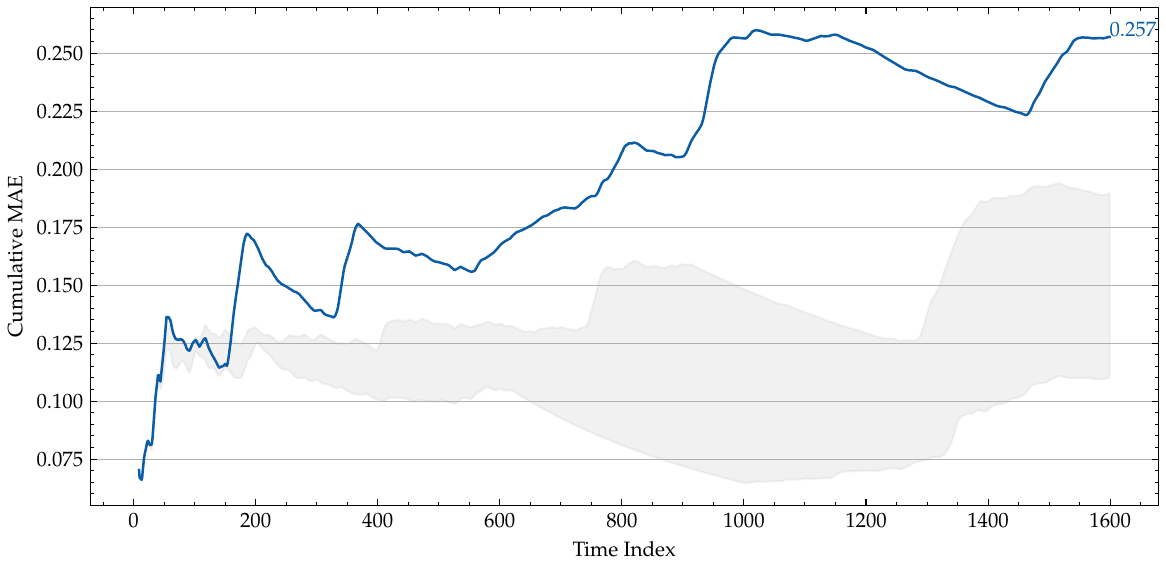}
        \caption{Static approach with $w=200$}
        \label{fig:mae_200}
    \end{subfigure}

    \caption{Cumulative MAE for the static approach with varying window sizes $w$. Shaded areas denote performance range across different window sizes.}
    \label{fig:dqsops_mae}
\end{figure*}

\begin{figure*}
    \centering
    \begin{subfigure}[b]{0.4\textwidth}
        \centering
        \includegraphics[width=\linewidth]{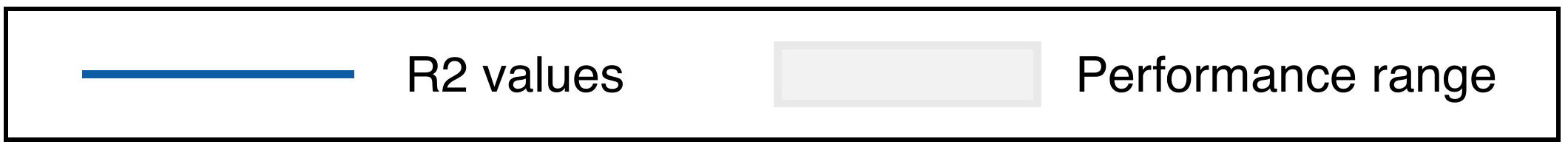}
    \end{subfigure}

    \begin{subfigure}[b]{0.48\textwidth}
        \centering
        \includegraphics[width=\linewidth]{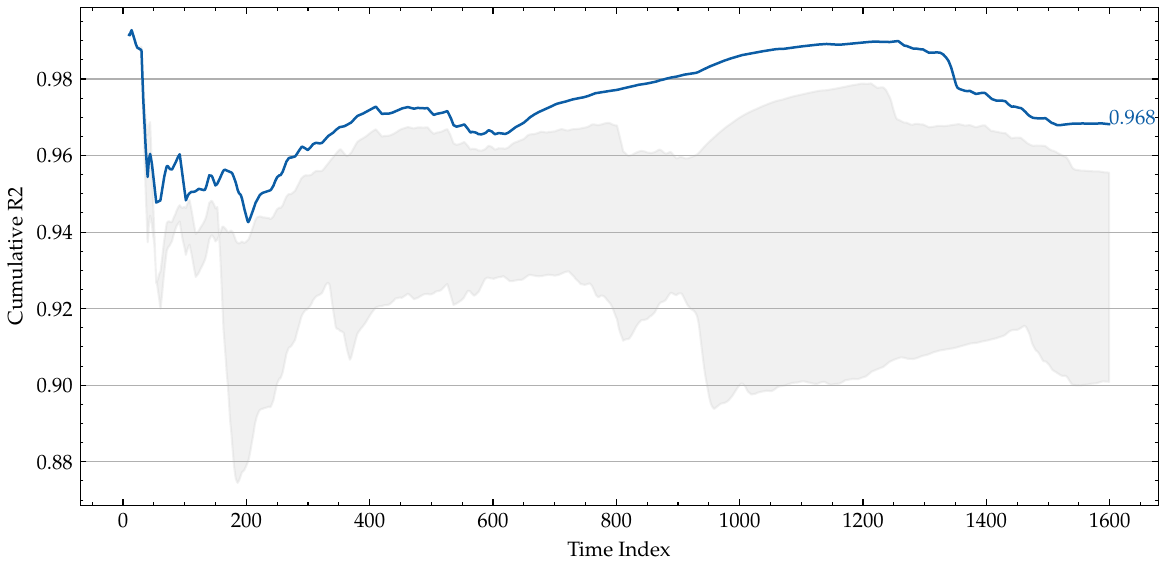}
        \caption{Static approach with $w=25$}
        \label{fig:r2_25}
    \end{subfigure}
    \hfill
    \begin{subfigure}[b]{0.48\textwidth}
        \centering
        \includegraphics[width=\linewidth]{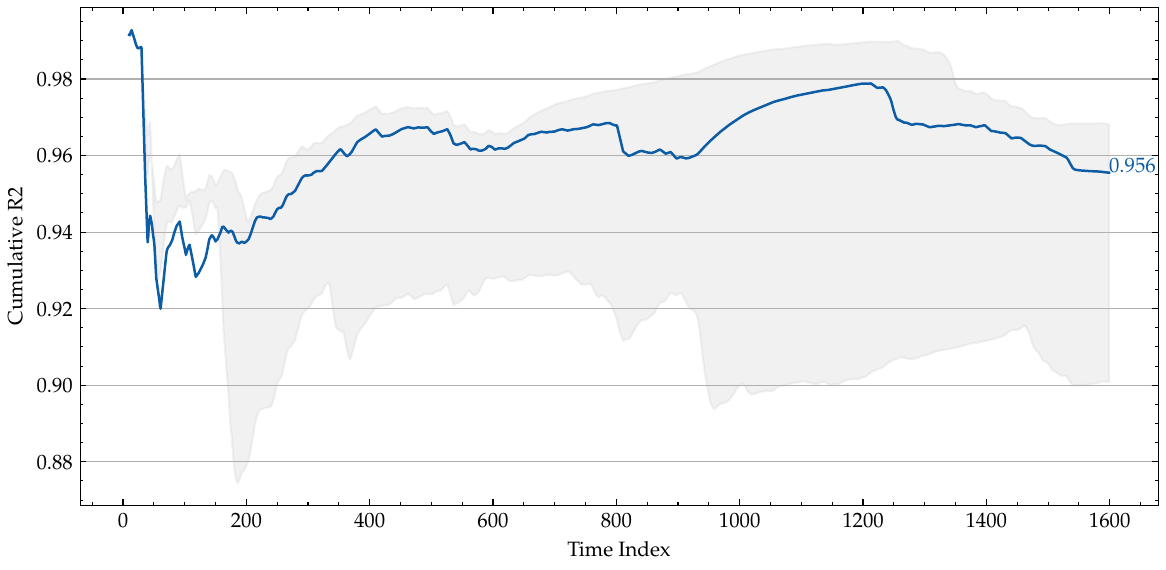}
        \caption{Static approach with $w=50$}
        \label{fig:r2_50}
    \end{subfigure}

    \vspace{10pt}

    \begin{subfigure}[b]{0.48\textwidth}
        \centering
        \includegraphics[width=\linewidth]{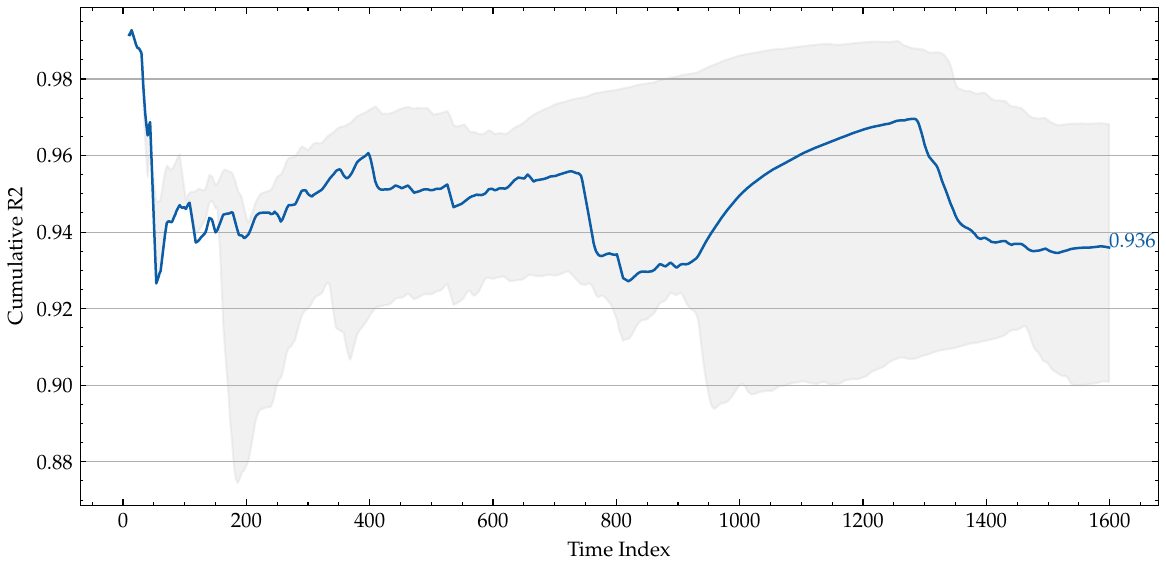}
        \caption{Static approach with $w=100$}
        \label{fig:r2_100}
    \end{subfigure}
    \hfill
    \begin{subfigure}[b]{0.48\textwidth}
        \centering
        \includegraphics[width=\linewidth]{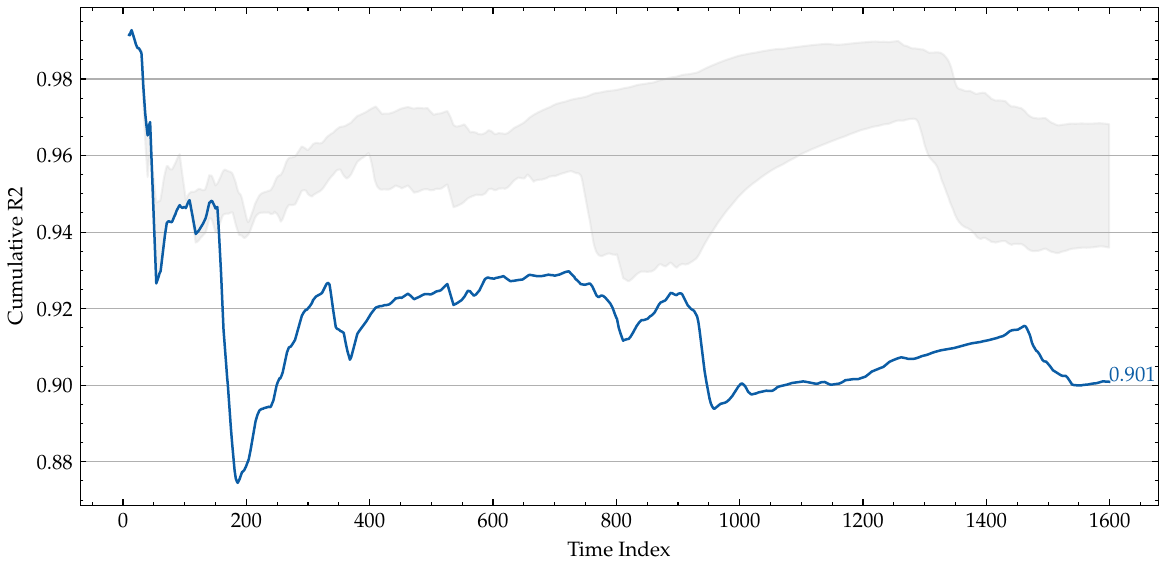}
        \caption{Static approach with $w=200$}
        \label{fig:r2_200}
    \end{subfigure}

    \caption{Cumulative R2 for the static approach with varying window sizes. Shaded areas denote performance range across different window sizes.}
    \label{fig:dqsops_r2}
\end{figure*}

We extend the analysis to evaluate the performance of the non-adaptive static approach, a framework dependent on a fixed window size $w$ for adaptation. Unlike the adaptive approach, where adaptation is determined by significance level $\tau$, the static approach triggers adaptation if the window size has been reached and the ML performance falls below a defined level. The experiments conducted with the static approach involve varying window sizes for adaptation triggers, with four different window sizes $w$: $25$, $50$, $100$, and $200$. Smaller window sizes represent a more frequent assessment for adaptations, while larger window sizes offer a wider temporal range. The corresponding MAE and $R^2$ metrics for the static approach are presented in Figures \ref{fig:dqsops_mae} and \ref{fig:dqsops_r2}, respectively. Each subfigure within these figures illustrates the cumulative error metrics over time for different window sizes. As with our adaptive approach, the shaded area represents the range of performance for static approaches with other window sizes.
 
The results show that smaller window sizes show superior performance in both the MAE and $R^2$ metrics. Specifically, for a window size of $25$, the MAE at the end of the experimental period is 0.11, while it is 0.257 for a window size of $200$. Furthermore, the metric $R^2$ for a window size of $25$ is 0.968, while it is 0.901 for a window size of $200$. Similarly to our adaptive approach, there is a noticeable improvement following the adaptation process. This improvement is particularly evident in the plot for window size 200, as depicted in Figures \ref{fig:mae_200} and \ref{fig:r2_200}. Specifically, these figures illustrate a significant performance drop before reaching the window size of $200$, indicating the need to update the predictor. In contrast, the window size of $25$ follows a more frequent adaptation pattern, resulting in a faster recovery from performance drops, as shown in Figures \ref{fig:mae_25} and \ref{fig:r2_25}.

When comparing our adaptive approach with the static approach, we can observe that both frameworks follow similar performance at the end of the experimental period. For the adaptive approach, optimal performance is observed at a significance threshold of $0.03$, which aligns the performance of the static approach with a window size of $25$. However, as we deviate from the optimal adaptation parameter of each methodology, performance starts to degrade gradually. Additionally, during the initial phases of the experiments, the performance of both adaptive and static approaches is more volatile, which can be explained by the models' sensitivity to changes in input, but as time progresses, the predictor increasingly matures and its performance becomes more consistent and stable. Meanwhile, the static approach displays a wider range of performance than the adaptive, reflecting the cumulative effect of errors across various window sizes.

\subsection{Time required analysis}

\begin{figure*}
    \centering
    \includegraphics[width=1\linewidth]{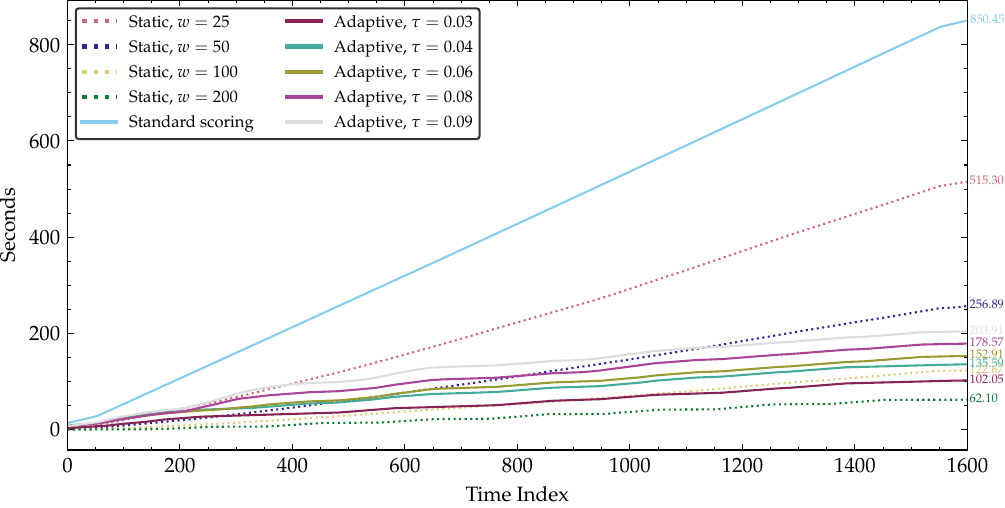}
    \caption{Cumulative time elapsed (in seconds) over time for the different scoring methodologies.}
    \label{fig:time_consumption}
\end{figure*}

The time required for various configurations was analyzed to understand their computational efficiency in the context of industrial tasks. Specifically, a comparison was performed between different scoring methodologies, including the adaptive, static, and the standard scoring approaches, to assess the time overhead in managing the data streams. The results of the time required to process the scoring task for the different approaches are summarized in Figure \ref{fig:time_consumption}.

Of all the approaches tested, we can see that the standard scoring methodology was the most time-consuming during the experimental period, requiring a total processing time of 850.45 seconds. However, the static approach with a window size of 200 achieved the shortest processing time among all methods due to the lower frequency of adaptation triggers associated with larger window sizes. We can also observe that both the standard scoring and static approaches exhibited linear trends, with processing times accelerating rapidly over time. In contrast, the adaptive approach showed a more conservative trend, with processing times increasing at a slower pace. The discrepancy is the result of the periodic adaptation mechanism inherent in the static approach, whereas the adaptive approach only adapts when drift is detected, leading to a more regulated processing speed.

Furthermore, the static approach demonstrated higher sensitivity to variations in the adaptation parameter, particularly the window size $w$, leading to increased dispersion in processing times. In particular, as we move from a larger $w$ of 200 to smaller sizes like 25, there is a substantial increase in processing time, from 62.15 seconds to 515.30 seconds. This sensitivity is also observed to a lesser extent in the adaptive approach. As indicated by the results, the adaptive approach with $\tau=0.03$ required a processing time of 102.05 seconds and reached 203.91 seconds for $\tau=0.09$. 

Overall, the time analysis reveals significant improvements in both processing efficiency and scalability with the adaptive approach. Specifically, the adaptive approach with $\tau=0.03$ achieved a processing time of 102.05 seconds—an 88\% reduction compared to the 850.45 seconds required by the standard scoring approach, representing a speedup factor of approximately 8.3x. In terms of scalability, Figure \ref{fig:time_consumption} illustrates that the processing time for the standard approach grows linearly with data volume, while our adaptive approach shows a more restrained growth rate. These trends suggest that the adaptive approach would maintain its efficiency advantage even as data volume scales.

\subsection{Analysis of Dynamic Data Quality Dimension Scores}
\begin{figure*}
    \centering
    \begin{subfigure}[b]{0.48\textwidth}
        \centering
        \includegraphics[width=\linewidth]{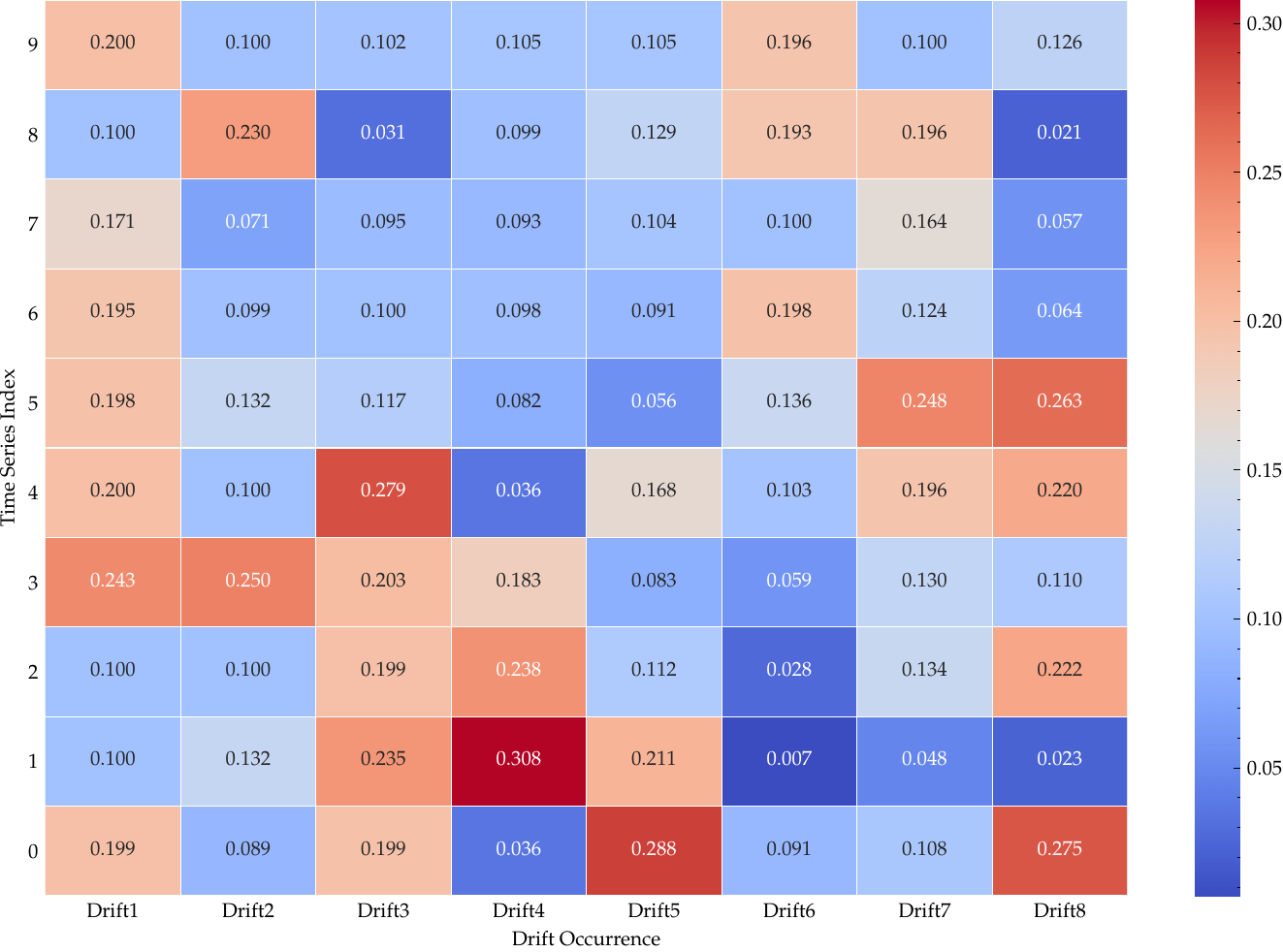}
        \caption{Change in skewness score after drift occurrences.}
        \label{fig:skewness}
    \end{subfigure}
    \hfill
    \begin{subfigure}[b]{0.48\textwidth}
        \centering
        \includegraphics[width=\linewidth]{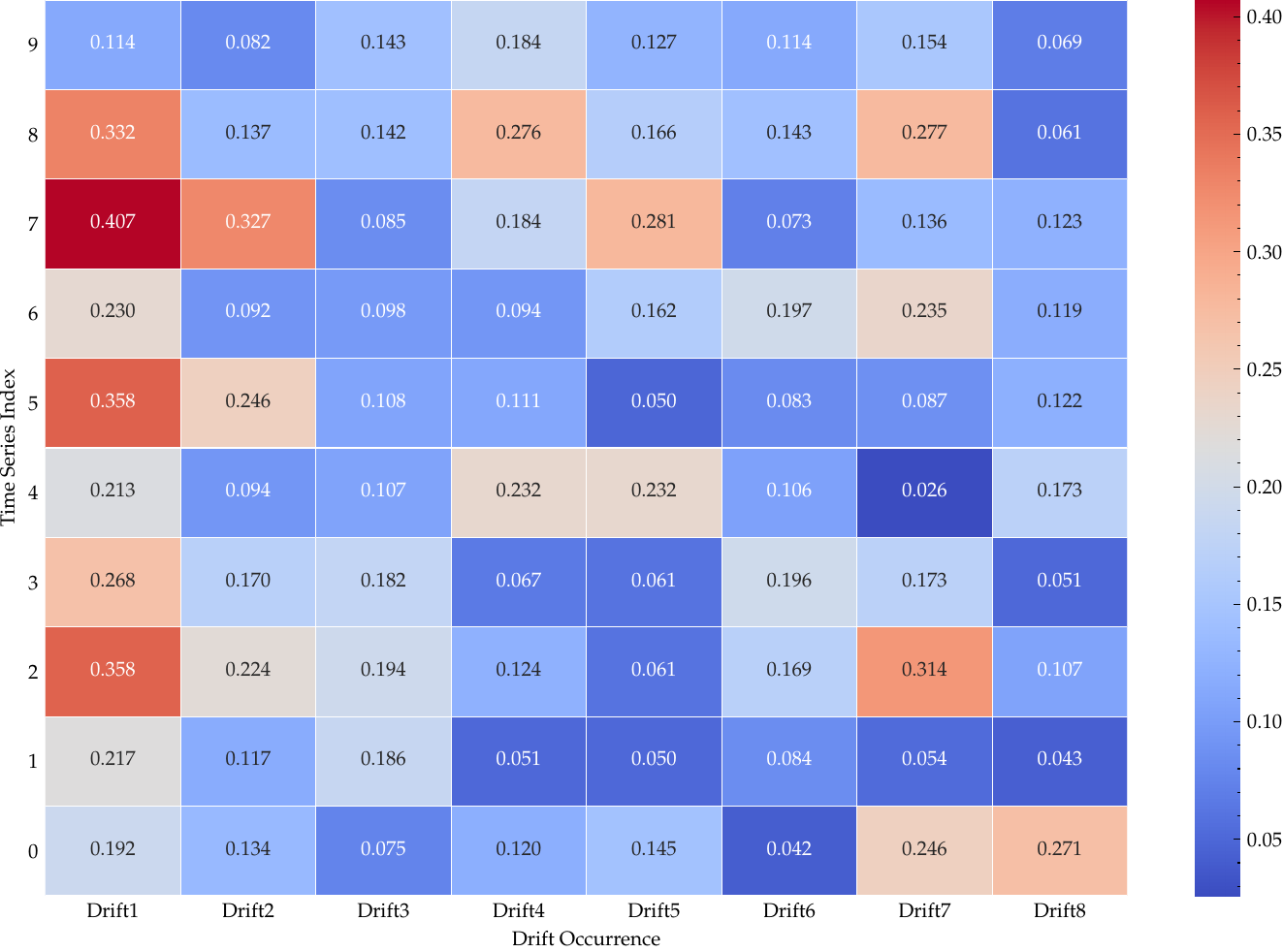}
        \caption{Change in timeliness score after drift occurrences.}
        \label{fig:timeliness}
    \end{subfigure}
    \caption{Difference in skewness and timeliness scores before and after drift occurrences.}
    \label{fig:sk_gof}
\end{figure*}

In this section, we analyze the evolution of dynamic data quality dimensions, specifically timeliness and skewness, under various drift occurrences. The analysis focuses on ten data points to highlight the dynamic nature of these data quality dimensions over time. The results of the discrepancy between each data point's score before and after each drift occurrence (for $\tau=0.03$ across 8 drift occurrences), indicating how much the score changed, are summarized in Figures \ref{fig:skewness} and \ref{fig:timeliness}, respectively. The numbers presented show the difference in the skewness and timeliness scores, highlighting how adaptation affects the score of the respective data quality dimension.

The findings reveal a substantial difference in magnitude after each adaptation, up to 0.3 for skewness and 0.4 for timeliness. The most significant change occurs immediately after the first occurrence after production, indicating that the system is more sensitive to change during this period. These experiments indicate a significant scale of change in the data quality dimension scores, signifying the impact of evolving patterns of the underlying data distribution. Furthermore, this shows the important role of adaptation in rescaling the data based on the prevalent situation, ensuring that the data quality dimensions accurately reflect the evolving characteristics of the dataset as their scores evolve over time.

\color{black}
\subsection{Resource Consumption}
\begin{figure}
    \centering
    \includegraphics[width=1\linewidth]{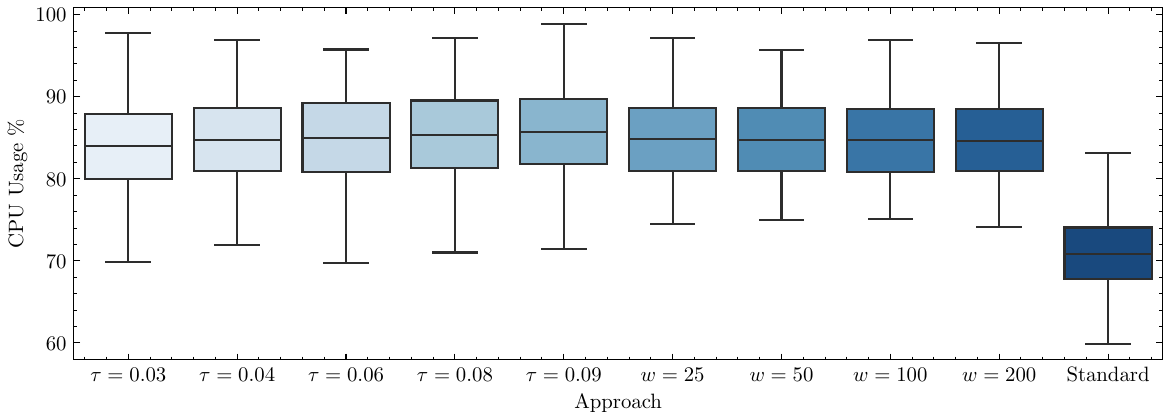}
    \caption{Summary of CPU usage percentage for the different scoring approaches}
    \label{fig:cpu_usage}
\end{figure}

\begin{figure}
    \centering
    \includegraphics[width=1\linewidth]{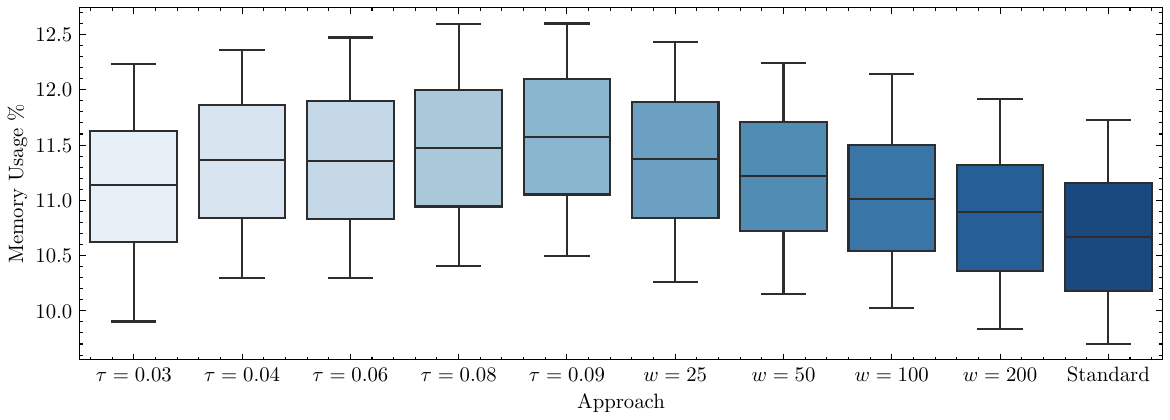}
    \caption{Summary of memory usage percentage for the different scoring approaches}
    \label{fig:memory_usage}
\end{figure}

The analysis of resource consumption explores the percentage of CPU and memory usage for each scoring approach, providing information on their respective resource demands in terms of computational and memory requirements. This analysis directly impacts operational costs and system efficiency, which are crucial factors in determining the most suitable approach to employ in production by industries. The percentage of CPU usage of the different approaches is summarized in Figure \ref{fig:cpu_usage}, while memory usage is summarized in Figure \ref{fig:memory_usage}.

The results indicate that the adaptive approach consumes slightly more CPU compared to the static, while the memory consumption between the two approaches is very similar. This difference in resource consumption can be attributed to the dynamic change detector mechanism employed by the adaptive approach, which actively monitors and updates the ML predictor based on detected drifts in the data streams, resulting in slightly higher computational overheads. On the contrary, the standard approach showed the lowest CPU and memory usage across the board due to its lack of ML involvement, resulting in a more straightforward and less resource-intensive process. Moreover, the analysis also reveals that variations in parameters within each approach do not significantly impact the memory consumption percentage, and the difference in the boxes is negligible. However, the distinction is more apparent when considering CPU usage, primarily due to the ML operations involved. Additionally, the adaptive approach tends to exhibit wider ranges between the CPU and memory usage whiskers compared to the static approach. This characteristic arises from the detection mechanism, which occasionally triggers adaptations in the ML predictor. These adaptations lead to higher fluctuations in resource usage over time.

\subsection{Operational Insights and Key Takeaways}
In this set of real-world industrial experiments, the behavior of data quality scoring approaches was analyzed in terms of prediction performance and processing time efficiency. In the following points, we analyze the main findings and lessons learned with an emphasis on practicality, aiming to offer valuable insights for industrial deployments:

\begin{enumerate}
    \item \textbf{Improved Predictive Performance:} We found that both the adaptive and static approaches show improvements in predictive performance after executing adaptation mechanisms. This improvement is reflected in the decrease in MAE and the increase in $R^2$ metrics post-adaptation. For adaptation parameters, smaller window sizes $w$ and lower significance thresholds $\tau$ resulted in higher performance improvements.
    \item  \textbf{Enhanced Processing Efficiency:} Regarding time efficiency, both frameworks demonstrated significant reductions in processing time compared to the standard data quality scoring approach. Comparing the two frameworks, we observed similar levels of predictive performance at the end of the experimental period using the optimal adaptation parameter. However, the adaptive approach achieved this performance with a more than two-fold improvement in processing time. This efficiency gain can be attributed to the fact that the adaptive approach triggers adaptations only when necessary, thus eliminating the costs associated with unnecessary adaptation processes. Moreover, both frameworks showed discernible trends in terms of acceleration and dispersion of processing time. The static approach showed a higher magnitude of acceleration in processing time. However, the adaptive approach displayed a more conservative trend, with processing times increasing at a steadier pace. This difference arises from the periodic adaptation mechanism inherent in the static approach, whereas the adaptive approach adapts only when drift is detected, resulting in a more regulated processing trendline. 
    \item \textbf{Sensitivity to Adaptation Parameters:} The static approach demonstrated a higher sensitivity to changes in the adaptation parameter, particularly the window size $w$, leading to a greater dispersion in processing times as this parameter varied. On the contrary, the adaptive approach showed less sensitivity to these changes, resulting in more consistent processing times over varying thresholds $\tau$. As the system progresses over time, this trend implies potential cost savings, serving as an indicator of the efficiency gained by integrating a drift detector to control the adaptation processes.
    \item \textbf{Minimal Impact on Resource Consumption:} The resource consumption analysis revealed that the adaptive approach utilizes slightly more CPU compared to the static approach, while the difference in memory usage between the two approaches is trivial. However, this additional CPU consumption is not significantly higher, implying that the differences in resource demands between the two approaches are relatively small. This indicates that the modifications executed in the adaptive approach do not add a significant burden on the data scoring system. Moreover, the standard approach was observed to be the least consuming in terms of CPU and memory resources, mainly because it does not involve ML components.
    \item \textbf{Key Industry Implications:} In our industrial context, maintaining high performance and ensuring timely processing of accurate data is crucial for critical use cases, such as preventing costly improper pumping events. The data quality scoring, as a component of a broader data-driven AI system, is crucial for making sound decisions. The analysis showed that the adaptive approach, with its dynamic nature, ensures that data quality scores remain up-to-date, reflecting the most recent state of the data streams. By activating adaptations only when required, the adaptive approach maintains current data scores and saves costs by avoiding unnecessary processing overheads. This adaptability is particularly advantageous in dynamic industrial environments where data characteristics may change over time, ensuring that the information derived from data quality is recent and reliable. Therefore, this dual benefit of up-to-date scores and cost savings demonstrates the benefits of the adaptive approach over other frameworks.
\end{enumerate}

\section{Conclusion}\label{sec:conclusion}
Assuring the quality of the incoming data streams is essential to build data-driven AI systems. An approach to quality assessment is data quality scoring, which allows quantification and evaluation of data quality. In this paper, we introduce the adaptive data quality scoring framework, which is designed to address the challenges of scoring dynamic data quality dimensions in industrial processes. The adaptive scoring framework is especially valuable in scenarios where scores should align with the current conditions of the system. The primary novelty of our framework is the use of a dynamic change detector, which controls the adaptation of the ML scoring predictor based on detected drifts in the data streams. Our proposed framework actively updates the ML predictor only when necessary, rather than passively updating it periodically, thus minimizing unnecessary computational overhead. Our experimental results demonstrate that implementing our framework leads to substantial enhancements in processing time efficiency with negligible impact on resource consumption and overall improved performance within a real-world use case. These enhancements are particularly reflected in the high predictive performance and reduced processing time, making our framework a feasible solution for critical industrial applications. Moving forward, we will investigate the integration of our framework into broader data-driven AI systems, where its real-time data quality scores can inform decision-making industrial processes by leveraging these quality scores to enhance the training of ML models on high-quality data.

\section*{Acknowledgement}
This work has been funded by the Knowledge Foundation of Sweden (KKS) through the Synergy Project AIDA - A Holistic AI driven Networking and Processing Framework for Industrial IoT
(Rek:20200067).



\bibliographystyle{ieeetr}

\bibliography{cas-refs}

\begin{thebibliography}{10}

\bibitem{reis2021data}
M.~S. Reis and P.~M. Saraiva, ``Data-centric process systems engineering: A push towards pse 4.0,'' {\em Computers \& Chemical Engineering}, vol.~155, p.~107529, 2021.

\bibitem{hazen2014data}
B.~T. Hazen, C.~A. Boone, J.~D. Ezell, and L.~A. Jones-Farmer, ``Data quality for data science, predictive analytics, and big data in supply chain management: An introduction to the problem and suggestions for research and applications,'' {\em International Journal of Production Economics}, vol.~154, pp.~72--80, 2014.

\bibitem{Stonebraker2019MachineLA}
M.~Stonebraker and E.~K. Rezig, ``Machine learning and big data: What is important?,'' {\em IEEE Data Eng. Bull.}, vol.~42, pp.~3--7, 2019.

\bibitem{zaveri2016quality}
A.~Zaveri, A.~Rula, A.~Maurino, R.~Pietrobon, J.~Lehmann, and S.~Auer, ``Quality assessment for linked data: A survey,'' {\em Semantic Web}, vol.~7, no.~1, pp.~63--93, 2016.

\bibitem{chen2014review}
H.~Chen, D.~Hailey, N.~Wang, and P.~Yu, ``A review of data quality assessment methods for public health information systems,'' {\em International journal of environmental research and public health}, vol.~11, no.~5, pp.~5170--5207, 2014.

\bibitem{liu2009encyclopedia}
L.~Liu and M.~T. {\"O}zsu, {\em Encyclopedia of database systems}, vol.~6.
\newblock Springer New York, NY, USA:, 2009.

\bibitem{batini2009methodologies}
C.~Batini, C.~Cappiello, C.~Francalanci, and A.~Maurino, ``Methodologies for data quality assessment and improvement,'' {\em ACM computing surveys (CSUR)}, vol.~41, no.~3, pp.~1--52, 2009.

\bibitem{taleb2021big}
I.~Taleb, M.~A. Serhani, C.~Bouhaddioui, and R.~Dssouli, ``Big data quality framework: a holistic approach to continuous quality management,'' {\em Journal of Big Data}, vol.~8, no.~1, pp.~1--41, 2021.

\bibitem{chen2021data}
H.~Chen, J.~Chen, and J.~Ding, ``Data evaluation and enhancement for quality improvement of machine learning,'' {\em IEEE Transactions on Reliability}, vol.~70, no.~2, pp.~831--847, 2021.

\bibitem{budach2022effects}
L.~Budach, M.~Feuerpfeil, N.~Ihde, A.~Nathansen, N.~Noack, H.~Patzlaff, F.~Naumann, and H.~Harmouch, ``The effects of data quality on machine learning performance,'' {\em arXiv preprint arXiv:2207.14529}, 2022.

\bibitem{widad2023quality}
E.~Widad, E.~Saida, and Y.~Gahi, ``Quality anomaly detection using predictive techniques: An extensive big data quality framework for reliable data analysis,'' {\em IEEE Access}, 2023.

\bibitem{wu2019real}
Z.~Wu, D.~Rincon, and P.~D. Christofides, ``Real-time adaptive machine-learning-based predictive control of nonlinear processes,'' {\em Industrial \& Engineering Chemistry Research}, vol.~59, no.~6, pp.~2275--2290, 2019.

\bibitem{bayram2023dqsops}
F.~Bayram, B.~S. Ahmed, E.~Hallin, and A.~Engman, ``{DQSOps}: Data quality scoring operations framework for data-driven applications,'' in {\em Proceedings of the 27th International Conference on Evaluation and Assessment in Software Engineering}, pp.~32--41, 2023.

\bibitem{clerc2016adaptive}
M.~Clerc, E.~Dauc{\'e}, and J.~Mattout, ``Adaptive methods in machine learning,'' {\em Brain--Computer Interfaces 1: Foundations and Methods}, pp.~207--232, 2016.

\bibitem{teh2020sensor}
H.~Y. Teh, A.~W. Kempa-Liehr, and K.~I.-K. Wang, ``Sensor data quality: A systematic review,'' {\em Journal of Big Data}, vol.~7, no.~1, pp.~1--49, 2020.

\bibitem{polyzotis2018data}
N.~Polyzotis, S.~Roy, S.~E. Whang, and M.~Zinkevich, ``Data lifecycle challenges in production machine learning: a survey,'' {\em ACM SIGMOD Record}, vol.~47, no.~2, pp.~17--28, 2018.

\bibitem{zha2023data}
D.~Zha, Z.~P. Bhat, K.-H. Lai, F.~Yang, and X.~Hu, ``Data-centric ai: Perspectives and challenges,'' in {\em Proceedings of the 2023 SIAM International Conference on Data Mining (SDM)}, pp.~945--948, SIAM, 2023.

\bibitem{agrahari2022concept}
S.~Agrahari and A.~K. Singh, ``Concept drift detection in data stream mining: A literature review,'' {\em Journal of King Saud University-Computer and Information Sciences}, vol.~34, no.~10, pp.~9523--9540, 2022.

\bibitem{ditzler2015learning}
G.~Ditzler, M.~Roveri, C.~Alippi, and R.~Polikar, ``Learning in nonstationary environments: A survey,'' {\em IEEE Computational Intelligence Magazine}, vol.~10, no.~4, pp.~12--25, 2015.

\bibitem{liu2013change}
S.~Liu, M.~Yamada, N.~Collier, and M.~Sugiyama, ``Change-point detection in time-series data by relative density-ratio estimation,'' {\em Neural Networks}, vol.~43, pp.~72--83, 2013.

\bibitem{micevska2021sddm}
S.~Micevska, A.~Awad, and S.~Sakr, ``Sddm: an interpretable statistical concept drift detection method for data streams,'' {\em Journal of intelligent information systems}, vol.~56, pp.~459--484, 2021.

\bibitem{kammerer2019anomaly}
K.~Kammerer, B.~Hoppenstedt, R.~Pryss, S.~St{\"o}kler, J.~Allgaier, and M.~Reichert, ``Anomaly detections for manufacturing systems based on sensor data—insights into two challenging real-world production settings,'' {\em Sensors}, vol.~19, no.~24, p.~5370, 2019.

\bibitem{liu2022concept}
A.~Liu, J.~Lu, Y.~Song, J.~Xuan, and G.~Zhang, ``Concept drift detection delay index,'' {\em IEEE Transactions on Knowledge and Data Engineering}, vol.~35, no.~5, pp.~4585--4597, 2022.

\bibitem{bayram2023lstm}
F.~Bayram, P.~Aupke, B.~S. Ahmed, A.~Kassler, A.~Theocharis, and J.~Forsman, ``Da-lstm: A dynamic drift-adaptive learning framework for interval load forecasting with lstm networks,'' {\em Engineering Applications of Artificial Intelligence}, vol.~123, p.~106480, 2023.

\bibitem{lin1991divergence}
J.~Lin, ``Divergence measures based on the shannon entropy,'' {\em IEEE Transactions on Information theory}, vol.~37, no.~1, pp.~145--151, 1991.

\bibitem{wang2023overview}
J.~Wang, Y.~Liu, P.~Li, Z.~Lin, S.~Sindakis, and S.~Aggarwal, ``Overview of﻿ data quality﻿: Examining the dimensions, antecedents, and impacts of data quality,'' {\em Journal of the Knowledge Economy}, pp.~1--20, 2023.

\bibitem{fan2022foundations}
W.~Fan and F.~Geerts, {\em Foundations of data quality management}.
\newblock Springer Nature, 2022.

\bibitem{pipino2002data}
L.~L. Pipino, Y.~W. Lee, and R.~Y. Wang, ``Data quality assessment,'' {\em Communications of the ACM}, vol.~45, no.~4, pp.~211--218, 2002.

\bibitem{wang1996beyond}
R.~Y. Wang and D.~M. Strong, ``Beyond accuracy: What data quality means to data consumers,'' {\em Journal of management information systems}, vol.~12, no.~4, pp.~5--33, 1996.

\bibitem{priestley2023survey}
M.~Priestley, F.~O’Donnell, and E.~Simperl, ``A survey of data quality requirements that matter in ml development pipelines,'' {\em ACM Journal of Data and Information Quality}, 2023.

\bibitem{karkouch2016data}
A.~Karkouch, H.~Mousannif, H.~Al~Moatassime, and T.~Noel, ``Data quality in internet of things: A state-of-the-art survey,'' {\em Journal of Network and Computer Applications}, vol.~73, pp.~57--81, 2016.

\bibitem{lewis2023electronic}
A.~E. Lewis, N.~Weiskopf, Z.~B. Abrams, R.~Foraker, A.~M. Lai, P.~R. Payne, and A.~Gupta, ``Electronic health record data quality assessment and tools: a systematic review,'' {\em Journal of the American Medical Informatics Association}, vol.~30, no.~10, pp.~1730--1740, 2023.

\bibitem{karkovskova2023data}
S.~Karko{\v{s}}kov{\'a}, ``Data governance model to enhance data quality in financial institutions,'' {\em Information Systems Management}, vol.~40, no.~1, pp.~90--110, 2023.

\bibitem{hasan2020current}
M.~M. Hasan, J.~Popp, and J.~Ol{\'a}h, ``Current landscape and influence of big data on finance,'' {\em Journal of Big Data}, vol.~7, no.~1, pp.~1--17, 2020.

\bibitem{wu2021comprehensive}
H.~Wu, A.~Lin, K.~C. Clarke, W.~Shi, A.~Cardenas-Tristan, and Z.~Tu, ``A comprehensive quality assessment framework for linear features from volunteered geographic information,'' {\em International Journal of Geographical Information Science}, vol.~35, no.~9, pp.~1826--1847, 2021.

\bibitem{mcgilvray2021executing}
D.~McGilvray, {\em Executing data quality projects: Ten steps to quality data and trusted information (TM)}.
\newblock Academic Press, 2021.

\bibitem{mansouri2023iot}
T.~Mansouri, M.~R. Sadeghi~Moghadam, F.~Monshizadeh, and A.~Zareravasan, ``Iot data quality issues and potential solutions: a literature review,'' {\em The Computer Journal}, vol.~66, no.~3, pp.~615--625, 2023.

\bibitem{fadlallah2023bigqa}
H.~Fadlallah, R.~Kilany, H.~Dhayne, R.~El~Haddad, R.~Haque, Y.~Taher, and A.~Jaber, ``Bigqa: Declarative big data quality assessment,'' {\em ACM Journal of Data and Information Quality}, vol.~15, no.~3, pp.~1--30, 2023.

\bibitem{chug2021statistical}
S.~Chug, P.~Kaushal, P.~Kumaraguru, and T.~Sethi, ``Statistical learning to operationalize a domain agnostic data quality scoring,'' {\em arXiv preprint arXiv:2108.08905}, 2021.

\bibitem{ardagna2018context}
D.~Ardagna, C.~Cappiello, W.~Sam{\'a}, and M.~Vitali, ``Context-aware data quality assessment for big data,'' {\em Future Generation Computer Systems}, vol.~89, pp.~548--562, 2018.

\bibitem{byabazaire2022end}
J.~Byabazaire, G.~M. O’Hare, and D.~T. Delaney, ``End-to-end data quality assessment using trust for data shared iot deployments,'' {\em IEEE Sensors Journal}, vol.~22, no.~20, pp.~19995--20009, 2022.

\bibitem{byabazaire2023iot}
J.~Byabazaire, G.~M. O’Hare, R.~Collier, and D.~Delaney, ``Iot data quality assessment framework using adaptive weighted estimation fusion,'' {\em Sensors}, vol.~23, no.~13, p.~5993, 2023.

\bibitem{evans2006scaling}
P.~Evans, ``Scaling and assessment of data quality,'' {\em Acta Crystallographica Section D: Biological Crystallography}, vol.~62, no.~1, pp.~72--82, 2006.

\bibitem{singh2015quality}
P.~Singh and B.~Suri, ``Quality assessment of data using statistical and machine learning methods,'' in {\em Computational Intelligence in Data Mining-Volume 2: Proceedings of the International Conference on CIDM, 20-21 December 2014}, pp.~89--97, Springer, 2015.

\bibitem{heinrich2018requirements}
B.~Heinrich, D.~Hristova, M.~Klier, A.~Schiller, and M.~Szubartowicz, ``Requirements for data quality metrics,'' {\em Journal of Data and Information Quality (JDIQ)}, vol.~9, no.~2, pp.~1--32, 2018.

\bibitem{lionis2021rssi}
A.~Lionis, K.~P. Peppas, H.~E. Nistazakis, and A.~Tsigopoulos, ``Rssi probability density functions comparison using jensen-shannon divergence and pearson distribution,'' {\em Technologies}, vol.~9, no.~2, p.~26, 2021.

\bibitem{ash2020warm}
J.~Ash and R.~P. Adams, ``On warm-starting neural network training,'' {\em Advances in neural information processing systems}, vol.~33, pp.~3884--3894, 2020.

\bibitem{kreuzberger2023machine}
D.~Kreuzberger, N.~K{\"u}hl, and S.~Hirschl, ``Machine learning operations (mlops): Overview, definition, and architecture,'' {\em IEEE Access}, 2023.

\bibitem{kumara2023requirements}
I.~Kumara, R.~Arts, D.~Di~Nucci, W.~J. Van Den~Heuvel, and D.~A. Tamburri, ``Requirements and reference architecture for mlops: Insights from industry,'' {\em Authorea Preprints}, 2023.

\bibitem{chen2016xgboost}
T.~Chen and C.~Guestrin, ``Xgboost: A scalable tree boosting system,'' in {\em Proceedings of the 22nd acm sigkdd international conference on knowledge discovery and data mining}, pp.~785--794, 2016.

\bibitem{kiangala2021effective}
S.~K. Kiangala and Z.~Wang, ``An effective adaptive customization framework for small manufacturing plants using extreme gradient boosting-xgboost and random forest ensemble learning algorithms in an industry 4.0 environment,'' {\em Machine Learning with Applications}, vol.~4, p.~100024, 2021.

\end{thebibliography}




\end{document}